\documentclass{article}

\usepackage{theorem}
\usepackage{latexsym}
\usepackage{amssymb}

\def\disp{\displaystyle}
\def\aa{{\alpha}}
\def\ep{{\varepsilon}}
\def\ra{\rightarrow}
\def\da{\downarrow}
\def\lora{\longrightarrow}
\def\lra{\leftrightarrow}
\def\qed{\rule{1mm}{2.5mm}}
\newcommand{\BZ}{{\Bbb Z}}

\theoremstyle{plain} \newtheorem{thm}{Theorem}[section]
\theoremstyle{plain} \newtheorem{prop}[thm]{Proposition}
\theoremstyle{plain} \newtheorem{lem}[thm]{Lemma}
\theoremstyle{plain} \newtheorem{rem}[thm]{Remark}
\theoremstyle{plain} \newtheorem{coro}[thm]{Corollary}
\theoremstyle{plain} \newtheorem{df}[thm]{Definition}

\makeatletter
 
 \@addtoreset{equation}{section}
\makeatother

\begin{document}

\title{On a class of algebraic solutions to Painlev\'e VI equation, \\
       its determinant formula and coalescence cascade}
\author{Tetsu Masuda \\
Department of Mathematics, Kobe University, \\
Rokko, Kobe, 657-8501, Japan \\
{\small masuda@math.kobe-u.ac.jp}}
\date{}

\maketitle

\begin{abstract}
A determinant formula for a class of algebraic solutions to 
Painlev\'e VI equation (P$_{\rm VI}$) is presented. 
This expression is regarded as a special case of the universal characters. 
The entries of the determinant are given by the Jacobi polynomials. 
Degeneration to the rational solutions of P$_{\rm V}$ and P$_{\rm III}$ 
is  discussed by applying the coalescence procedure. 
Relationship between Umemura polynomials associated with P$_{\rm VI}$ 
and our formula is also discussed. 
\end{abstract}

\section{Introduction \label{intr}}
Enlarging the work by Yablonskii and Vorob'ev for P$_{\rm II}$~\cite{YV}
and Okamoto for P$_{\rm IV}$~\cite{O3}, 
Umemura has introduced a class of special polynomials associated with
the algebraic solutions of each Painlev\'e equation P$_{\rm III}$,
P$_{\rm V}$ and P$_{\rm VI}$~\cite{Um}. 
These polynomials are generated by Toda equation that arises from
B\"acklund transformations of each Painlev\'e equation. 
It is also known that the coefficients of the polynomials admit
mysterious combinatorial properties~\cite{NOOU,Tane}. 

It is remarkable that some of such polynomials are expressed as a
specialization of the Schur functions. 
Yablonskii-Vorob'ev polynomials are expressible by 2-reduced Schur
functions, 
and Okamoto polynomials by 3-reduced Schur functions~\cite{P2:rat,P4:rat,NY:P4}. 
It is now recognized that these structures reflect the affine Weyl group
symmetry as groups of the B\"acklund transformations~\cite{Ya}. 
The determinant formulas of Jacobi-Trudi type for Umemura polynomials of
P$_{\rm III}$ and P$_{\rm V}$ resemble each other. 
In both cases, they are expressed by 2-reduced Schur functions and
entries of the determinant are given by the Laguerre polynomials~\cite{P3:rat,NY:P5}. 

Furthermore, in a recent work, 
it has been revealed that the whole families of the characteristic
polynomials for the rational solutions of P$_{\rm V}$, 
which include Umemura polynomials for P$_{\rm V}$ as a special case, 
admit more general structures~\cite{P5:rat}. 
Namely, they are expressed in terms of the universal characters that are
a generalization of the Schur functions. 
The latter are the characters of the irreducible polynomial
representations of $GL(n)$, 
while the former are introduced to describe the irreducible rational
representations~\cite{Koike}. 

What kind of determinant structures do Umemura polynomials for 
P$_{\rm VI}$ admit? 
Recently, Kirillov and Taneda have introduced a generalization of
Umemura polynomials for P$_{\rm VI}$ in the context of combinatorics and
shown that their polynomials degenerate to the special polynomials for
P$_{\rm V}$ in some limit~\cite{KT1,KT2,KT3}. 
This result suggests that the special polynomials associated with the
algebraic solutions of P$_{\rm VI}$ are also expressible by the
universal characters. 

In this paper, we consider P$_{\rm VI}$ 
\begin{equation}
\begin{array}{l}
\disp
 \frac{d^2y}{dt^2}=
 \frac{1}{2}\left(\frac{1}{y}+\frac{1}{y-1}+\frac{1}{y-t}\right)\left(\frac{dy}{dt}\right)^2
 -\left(\frac{1}{t}+\frac{1}{t-1}+\frac{1}{y-t}\right)\frac{dy}{dt} \\ \\
\disp
 \hskip50pt+\frac{y(y-1)(y-t)}{2t^2(t-1)^2}
 \left[\kappa_{\infty}^2-\kappa_0^2\frac{t}{y^2}
      +\kappa_1^2\frac{t-1}{(y-1)^2}+(1-\theta^2)\frac{t(t-1)}{(y-t)^2}\right],
\end{array}   \label{P6}
\end{equation}
where $\kappa_{\infty}$, $\kappa_0$, $\kappa_1$ and $\theta$ are
parameters. 
As is well known~\cite{O1}, 
P$_{\rm VI}$ (\ref{P6}) is equivalent to the Hamilton system
\begin{equation}
\hskip-40pt
\mbox{S$_{\rm VI}$~:} \hskip30pt
q'=\frac{\partial H}{\partial p}, \quad p'=-\frac{\partial H}{\partial q}, \quad
'=t(t-1)\frac{d}{dt}, \label{cano6}
\end{equation}
with the Hamiltonian
\begin{equation}
\begin{array}{c}
\medskip
\disp
H=q(q-1)(q-t)p^2-[\kappa_0 (q-1)(q-t)+\kappa_1 q(q-t)+(\theta-1)q(q-1)]p+\kappa(q-t), \\
\disp
\kappa=\frac{1}{4}(\kappa_0+\kappa_1+\theta-1)^2-\frac{1}{4}\kappa_{\infty}^2. 
\end{array}   \label{H6}
\end{equation}
In fact, the equation for $y=q$ is nothing but P$_{\rm VI}$ (\ref{P6}). 

The aim of this paper is to investigate a class of algebraic solutions
to P$_{\rm VI}$ (or S$_{\rm VI}$) that originate from the fixed points
of the B\"acklund transformations corresponding Dynkin automorphisms 
and to present its explicit determinant formula. 

Let us remark on the terminology of ``algebraic solutions''. 
P$_{\rm VI}$ admits the several classes of algebraic solutions~\cite{Pi,DM,Ma,AK}, 
and the classification has not established. 
In this paper, if we do not comment especially, 
we use ``algebraic solutions'' in the above restricted sense. 

This paper is organized as follows. 
In Section \ref{main}, 
we present a determinant formula for the algebraic solutions of 
P$_{\rm VI}$ (or S$_{\rm VI}$). 
This expression is also a specialization of the universal characters 
and the entries of the determinant are given by the Jacobi polynomials. 
The symmetry of P$_{\rm VI}$ is described by the affine Weyl group of
type $D_4^{(1)}$. 
We give a symmetric description of B\"acklund transformations of 
P$_{\rm VI}$ and derive several sets of bilinear equations for the
$\tau$-functions in Section \ref{sym}. 
In Section \ref{constr}, 
we construct a family of algebraic solutions of P$_{\rm VI}$. 
Proof of our result is given in Section \ref{proof}. 
As is well known, 
P$_{\rm VI}$ degenerates to P$_{\rm V}, \ldots$, P$_{\rm I}$ by
successive limiting procedures~\cite{Pa,GtoP}. 
In Section \ref{degen}, 
we show that the algebraic solutions of P$_{\rm VI}$ degenerate to the
rational solutions of P$_{\rm V}$ and P$_{\rm III}$ with preserving the
determinant structures. 
Section \ref{rel:Ume} is devoted to discuss the relationship to the
original Umemura polynomials for P$_{\rm VI}$.

\section{A determinant formula \label{main}}
\begin{df}\label{def:V}
Let $p_k=p_k^{(c,d)}(x)$ and $q_k=q_k^{(c,d)}(x),~k \in \BZ$, be two 
sets of polynomials defined by 
\begin{equation}
\begin{array}{c}
\disp 
\sum_{k=0}^{\infty}p_k^{(c,d)}(x)\lambda^k=G(x;c,d;\lambda), \quad 
p_k^{(c,d)}(x)=0\ \mbox{for}\ k<0, \\
\disp 
q_k^{(c,d)}(x)=p_k^{(c,d)}(x^{-1}), 
\end{array}   \label{def:pq:V}
\end{equation}
respectively, where the generating function $G(x;c,d;\lambda)$ is given by 
\begin{equation}
G(x;c,d;\lambda)=(1-\lambda)^{c-d}\left(1+x\lambda\right)^{-c}. \label{GF:V}
\end{equation}
For $m,n \in \BZ_{\ge 0}$, we define a family of polynomials 
$R_{m,n}=R_{m,n}(x;c,d)$ by 
\begin{equation}
R_{m,n}(x;c,d)=
 \left|
  \begin{array}{cccccccc}
   q_1        & q_0        & \cdots     & q_{-m+2}   &
   q_{-m+1}   & \cdots     & q_{-m-n+3} & q_{-m-n+2} \\
   q_3        & q_2        & \cdots     & q_{-m+4}   &
   q_{-m+3}   & \cdots     & q_{-m-n+5} & q_{-m-n+4} \\
   \vdots     & \vdots     & \ddots     & \vdots     &
   \vdots     & \ddots     & \vdots     & \vdots     \\
   q_{2m-1}   & q_{2m-2}   & \cdots     & q_m        &
   q_{m-1}    & \cdots     & q_{m-n+1}  & q_{m-n}    \\
   p_{n-m}    & p_{n-m+1}  & \cdots     & p_{n-1}    &
   p_n        & \cdots     & p_{2n-2}   & p_{2n-1}   \\
   \vdots     & \vdots     & \ddots     & \vdots     &
   \vdots     & \ddots     & \vdots     & \vdots     \\
   p_{-n-m+4} & p_{-n-m+5} & \cdots     & p_{-n+3}   &
   p_{-n+4}   & \cdots     & p_2        & p_3        \\
   p_{-n-m+2} & p_{-n-m+3} & \cdots     & p_{-n+1}   &
   p_{-n+2}   & \cdots     & p_0        & p_1
  \end{array}
 \right|.    \label{P6:alg:tau}
\end{equation}
For $m,n \in \BZ_{<0}$, $R_{m,n}$ are defined through 
\begin{equation}
R_{m,n}=(-1)^{m(m+1)/2}R_{-m-1,n}, \quad 
R_{m,n}=(-1)^{n(n+1)/2}R_{m,-n-1}. \label{neg:R}
\end{equation}
\end{df}

\begin{thm}\label{detV}
We set 
\begin{equation}
R_{m,n}(x;c,d)=S_{m,n}(x;a,b), \label{S-R:V}
\end{equation}
with 
\begin{equation}
c=a+b+n-\frac{1}{2}, \quad d=2b-m+n. \label{abcd:V}
\end{equation}
Then, for the parameters 
\begin{equation}
\kappa_{\infty}=b, \quad \kappa_0=b-m+n, \quad \kappa_1=a+m+n, \quad \theta=a, 
\label{para}
\end{equation}
we have a family of algebraic solutions of the Hamilton system S$_{\rm VI}$, 
\begin{equation}
\begin{array}{c}
\medskip
\disp 
q=x\frac{S_{m,n-1}(x;a+1,b)S_{m-1,n}(x;a+1,b)}
        {S_{m-1,n}(x;a+1,b-1)S_{m,n-1}(x;a+1,b+1)}, \\ 
\disp 
p=\frac{1}{2}\left(a+b+n-\frac{1}{2}\right)x^{-1}
  \frac{S_{m-1,n}(x;a+1,b-1)S_{m,n-1}(x;a+1,b+1)S_{m,n-1}(x;a,b)}
       {S_{m,n}(x;a,b)S_{m-1,n-1}(x;a+1,b)S_{m,n-1}(x;a+1,b)}, 
\end{array}   \label{f-S:V}
\end{equation}
with $x^2=t$. 
\end{thm}

This Theorem means that a class of algebraic solutions of P$_{\rm VI}$ 
is expressed in terms of the universal characters~\cite{Koike}, 
which also appear in the expression of the rational solutions of 
P$_{\rm V}$~\cite{P5:rat}. 
Note that the entries $p_k$ and $q_k$ are essentially the Jacobi
polynomials, namely, 
\begin{equation}
p_k^{(c,d)}(x)=P_k^{(d-1,c-d-k)}(-1-2x). 
\end{equation}

Applying some B\"acklund transformations, which can include outer
transformations given in (\ref{outer}), to the above solutions, 
we can get other families of algebraic solutions of P$_{\rm VI}$. 
Some examples are presented in Corollary \ref{toV} and \ref{toIII}, 
so we omit detail here.

\section{A symmetric description of Painlev\'e VI equation \label{sym}}
Noumi and Yamada have introduced the symmetric form of Painlev\'e
equations~\cite{NY1,NY2,NY:P4}. 
This formulation provides us with a clear description of symmetry
structures of B\"acklund transformations and a systematic tool to
construct special solutions. 

In this section, we present a symmetric description for the B\"acklund
 transformations of P$_{\rm VI}$~\cite{NY3,private}. 
After introducing the $\tau$-functions via Hamiltonians, 
we derive several sets of bilinear equations.

\subsection{B\"acklund transformations of P$_{\rm VI}$}
We set 
\begin{equation}
f_0=q-t, \quad f_3=q-1, \quad f_4=q, \quad f_2=p,
\end{equation}
and 
\begin{equation}
\aa_0=\theta, \quad \aa_1=\kappa_{\infty}, \quad \aa_3=\kappa_1, \quad \aa_4=\kappa_0. 
\end{equation}
Then, the Hamiltonian (\ref{H6}) is written as 
\begin{equation}
H=f_2^2f_0f_3f_4-[(\aa_0-1)f_3f_4+\aa_3 f_0f_4+\aa_4 f_0f_3]f_2+\aa_2(\aa_1+\aa_2)f_0,  
\label{H6:f}
\end{equation}
with 
\begin{equation}
\aa_0+\aa_1+2\aa_2+\aa_3+\aa_4=1, 
\end{equation}
and the Hamilton equation (\ref{cano6}) is written down as 
\begin{equation}
\begin{array}{l}
\smallskip
\disp
f_4'=2f_2f_0f_3f_4-(\aa_0-1)f_3f_4-\aa_3 f_0f_4-\aa_4 f_0f_3, \\
\smallskip
\disp
f_2'=-(f_0f_3+f_0f_4+f_3f_4)f_2^2 \\
\disp
\hskip40pt
     +[(\aa_0-1)(f_3+f_4)+\aa_3(f_0+f_4)+\aa_4(f_0+f_3)]f_2-\aa_2(\aa_1+\aa_2). 
\end{array}   \label{cano6'}
\end{equation}

The B\"acklund transformations of P$_{\rm VI}$ are described as follows~\cite{KMNOY,NY3},
\begin{equation}
s_i(\aa_j)=\aa_j-a_{ij}\aa_i, \quad (i,j=0,1,2,3,4)  \label{BT:a}
\end{equation}
\begin{equation}
s_2(f_i)=f_i+\frac{\aa_2}{f_2}, \quad s_i(f_2)=f_2-\frac{\aa_i}{f_i}, \quad (i=0,3,4) 
\label{BT:f}
\end{equation}
\begin{eqnarray}
s_5: 
&&\aa_0 \lra \aa_1, \quad \aa_3 \lra \aa_4, \nonumber \\
&&f_2 \ra -\frac{f_0(f_2f_0+\aa_2)}{t(t-1)}, \quad 
  f_0 \ra \frac{t(t-1)}{f_0}, \quad f_3 \ra (t-1)\frac{f_4}{f_0}, \quad f_4 \ra t\frac{f_3}{f_0},\\
s_6:
&&\aa_0 \lra \aa_3, \quad \aa_1 \lra \aa_4, \nonumber \\
&&f_2 \ra -\frac{f_4(f_4f_2+\aa_2)}{t},\quad 
  f_0 \ra -t\frac{f_3}{f_4}, \quad f_3 \ra -\frac{f_0}{f_4}, \quad f_4 \ra \frac{t}{f_4}, \label{s6} \\
s_7:
&&\aa_0 \lra \aa_4,\quad \aa_1 \lra \aa_3, \nonumber \\
&&f_2 \ra \frac{f_3(f_3f_2+\aa_2)}{t-1}, \quad
  f_0 \ra -(t-1)\frac{f_4}{f_3}, \quad f_3 \ra -\frac{t-1}{f_3}, \quad f_4 \ra \frac{f_0}{f_3},
\end{eqnarray}
where $A=(a_{ij})_{i,j=0}^4$ is the Cartan matrix of type $D^{(1)}_4$: 
\begin{equation}
a_{ii}=2 \quad (i=0,1,2,3,4), \quad 
a_{2j}=a_{j2}=-1, \quad (j=0,1,3,4), \quad
a_{ij}=0 \quad (\mbox{otherwise}).
\end{equation}
These transformations commute with the derivation $'$, and satisfy the following relations
\begin{equation}
\begin{array}{c}
\disp
  s_i^2=1 \quad (i=0,\ldots,7), \quad s_i s_2 s_i=s_2 s_i s_2 \quad (i=0,1,3,4), \\
\disp
  s_5 s_{\{0,1,2,3,4\}}=s_{\{1,0,2,4,3\}} s_5, \quad 
  s_6 s_{\{0,1,2,3,4\}}=s_{\{3,4,2,0,1\}} s_6, \quad 
  s_7 s_{\{0,1,2,3,4\}}=s_{\{4,3,2,1,0\}} s_7, \\
\disp
  s_5 s_6=s_6 s_5, \quad s_5 s_7=s_7 s_5, \quad s_6 s_7=s_7 s_6. 
\end{array}
\end{equation}
This means that transformations $s_i~(i=0,\dots,4)$ generate the affine Weyl group $W(D^{(1)}_4)$, and $s_i~(i=0,\dots,7)$ generate its extension including the Dynkin diagram automorphisms.

\subsection{The $\tau$-functions and bilinear equations}
We add a correction term to the Hamiltonian (\ref{H6:f}) as follows, 
\begin{equation}
H_0=H+
\frac{t}{4}\left[1+4\aa_1\aa_2+4\aa_2^2-(\aa_3+\aa_4)^2\right]
+\frac{1}{4}\left[(\aa_1+\aa_4)^2+(\aa_3+\aa_4)^2+4\aa_2\aa_4\right].   \label{corr:H}
\end{equation}
This modification gives a simpler behavior of the Hamiltonian with respect
to the B\"acklund transformations. 
From the corrected Hamiltonian (\ref{corr:H}), we introduce a family of
Hamiltonians $h_i(i=0,1,2,3,4)$ as 
\begin{equation}
h_0=H_0+\frac{t}{4},\quad h_1=s_5(H_0)-\frac{t-1}{4},\quad 
h_3=s_6(H_0)+\frac{1}{4},\quad h_4=s_7(H_0), \quad h_2=h_1+s_1(h_1).  \label{h:def}
\end{equation}
Then, we have 
\begin{equation}
s_i(h_j)=h_j, \quad (i \ne j,~i,j=0,1,2,3,4), \label{BT:h:trvl}
\end{equation}
\begin{equation}
\begin{array}{c}
\smallskip
\disp 
s_0(h_0)=h_0-\aa_0(t-1)\frac{f_4}{f_0}, \quad s_1(h_1)=h_1-\aa_1 f_3, \\
\disp 
s_3(h_3)=h_3+\aa_3\frac{t-1}{f_3}, \quad s_4(h_4)=h_4+\aa_4\frac{f_0}{f_4}. 
\end{array}   \label{BT:h}
\end{equation}
Moreover, from (\ref{h:def}),(\ref{BT:h}) and the equations (\ref{cano6'}), we obtain 
\begin{equation}
\begin{array}{c}
\disp
[s_i(h_i)+h_i]-[s_1(h_1)+h_1]=\frac{f_i'}{f_i}, \quad (i=0,3,4) \\
\disp
[s_2(h_2)+h_2]-(h_0+h_1+h_3+h_4)=\frac{f_2'}{f_2}-\frac{1}{2}(t-1).
\end{array}   \label{BT:logtau}
\end{equation}

Next, we also introduce $\tau$-functions $\tau_i~(i=0,1,2,3,4)$ by 
\begin{equation}
h_i=\frac{\tau_i'}{\tau_i}.
\end{equation}
Implying that the action of $s_i$'s on $\tau$-functions also commute
with the derivation $'$, 
one can lift the B\"acklund transformations to the $\tau$-functions. 
From (\ref{BT:h:trvl}) and (\ref{BT:logtau}), we get 
\begin{equation}
s_i(\tau_j)=\tau_j, \quad (i\neq j,~i,j=0,1,2,3,4), 
\end{equation}
and 
\begin{equation}
s_0(\tau_0)=f_0\frac{\tau_2}{\tau_0}, \quad 
s_1(\tau_1)=\frac{\tau_2}{\tau_1}, \quad 
s_2(\tau_2)=t^{-\frac{1}{2}}f_2\frac{\tau_0\tau_1\tau_3\tau_4}{\tau_2}, \quad 
s_3(\tau_3)=f_3\frac{\tau_2}{\tau_3}, \quad 
s_4(\tau_4)=f_4\frac{\tau_2}{\tau_4},      \label{BT:tau}
\end{equation}
respectively. 
The action of the diagram automorphisms $s_5,s_6$ and $s_7$ are derived
by using (\ref{h:def}) as follows, 
\begin{eqnarray}
s_5:
&& \quad \tau_0 \ra [t(t-1)]^{\frac{1}{4}}\tau_1, 
   \quad \tau_1 \ra [t(t-1)]^{-\frac{1}{4}}\tau_0,             \cr
&& \quad \tau_3 \ra t^{-\frac{1}{4}}(t-1)^{\frac{1}{4}}\tau_4, 
   \quad \tau_4 \ra t^{\frac{1}{4}}(t-1)^{-\frac{1}{4}}\tau_3, \\
&& \quad \tau_2 \ra [t(t-1)]^{-\frac{1}{2}}f_0\tau_2,          \nonumber
\end{eqnarray}
\begin{equation}
s_6: \quad \tau_0 \ra i t^{\frac{1}{4}}\tau_3, 
     \quad \tau_3 \ra -i t^{-\frac{1}{4}}\tau_0, 
     \quad \tau_1 \ra t^{-\frac{1}{4}}\tau_4,  
     \quad \tau_4 \ra t^{\frac{1}{4}}\tau_1, 
     \quad \tau_2 \ra t^{-\frac{1}{2}}f_4 \tau_2,  \label{s6totau}
\end{equation}
\begin{eqnarray}
s_7:
&& \quad \tau_0 \ra (-1)^{-\frac{3}{4}}(t-1)^{\frac{1}{4}}\tau_4, 
   \quad \tau_4 \ra (-1)^{\frac{3}{4}}(t-1)^{-\frac{1}{4}}\tau_0,\cr
&& \quad \tau_1 \ra (-1)^{\frac{3}{4}}(t-1)^{-\frac{1}{4}}\tau_3, 
   \quad \tau_3 \ra (-1)^{-\frac{3}{4}}(t-1)^{\frac{1}{4}}\tau_1,\\
&& \quad \tau_2 \ra -i(t-1)^{-\frac{1}{2}}f_3\tau_2.             \nonumber
\end{eqnarray}
The algebraic relations of $s_i$'s are preserved in this lifting 
except for the following modification,
\begin{equation}
s_i s_2 (\tau_2)=-s_2 s_i (\tau_2) \quad (i=5,6,7), \label{modi:1}
\end{equation}
and 
\begin{equation}
\begin{array}{l}
\disp
 s_5 s_6 \tau_{\{0,1,2,3,4\}}=\{i,-i,-1,-i,i\} s_6 s_5 \tau_{\{0,1,2,3,4\}}, \\
\disp
 s_5 s_7 \tau_{\{0,1,2,3,4\}}=\{i,-i,-1,i,-i\} s_7 s_5 \tau_{\{0,1,2,3,4\}}, \\
\disp
 s_6 s_7 \tau_{\{0,1,2,3,4\}}=\{-i,-i,-1,i,i\} s_7 s_6 \tau_{\{0,1,2,3,4\}}.
\end{array}   \label{modi:2}
\end{equation}
Note that one can regard (\ref{BT:tau}) as the multiplicative
formulas for $f_i$ in terms of $\tau$-functions, 
\begin{equation}
f_0=\frac{\tau_0s_0(\tau_0)}{\tau_1s_1(\tau_1)}, \quad 
f_3=\frac{\tau_3s_3(\tau_3)}{\tau_1s_1(\tau_1)}, \quad 
f_4=\frac{\tau_4s_4(\tau_4)}{\tau_1s_1(\tau_1)}, \quad 
f_2=t^{\frac{1}{2}}\frac{\tau_1s_1(\tau_1)s_2s_1(\tau_1)}{\tau_0\tau_3\tau_4}. 
\label{multi-form}
\end{equation}
From these formulas, it is possible to derive various bilinear equations
for $\tau$-functions. 
First, the constraints for $f$-variables 
\begin{equation}
f_0=f_4-t, \quad f_3=f_4-1, 
\end{equation}
yield to 
\begin{equation}
\begin{array}{c}
\smallskip
\disp
\tau_1 s_1(\tau_1)+\tau_3 s_3(\tau_3)-\tau_4 s_4(\tau_4)=0,  \\
\smallskip
\disp
\tau_0 s_0(\tau_0)+t\tau_1 s_1(\tau_1)-\tau_4 s_4(\tau_4)=0,  \\
\smallskip
\disp
\tau_1 s_2s_1(\tau_1)+ \tau_3 s_2s_3(\tau_3)-\tau_4 s_2s_4(\tau_4)=0, \\
\disp
\tau_0 s_2s_0(\tau_0)+t\tau_1 s_2s_1(\tau_1)-\tau_4 s_2s_4(\tau_4)=0. 
\end{array}   \label{bi:norm}
\end{equation}
The B\"acklund transformations (\ref{BT:f}) are lead to the following
sets of bilinear equations, 
\begin{equation}
\begin{array}{c}
\smallskip
\disp
\aa_0 t^{-\frac{1}{2}}     \tau_3\tau_4-s_0(\tau_0)s_2s_1(\tau_1)+\tau_0 s_0s_2s_1(\tau_1)=0, \\
\smallskip
\disp
\aa_0 t^{-\frac{1}{2}}(t-1)\tau_1\tau_4-s_0(\tau_0)s_2s_3(\tau_3)+\tau_0 s_0s_2s_3(\tau_3)=0, \\
\smallskip
\disp
\aa_0 t^{\frac{1}{2}}      \tau_1\tau_3-s_0(\tau_0)s_2s_4(\tau_4)+\tau_0 s_0s_2s_4(\tau_4)=0,
\end{array}
\end{equation}
\begin{equation}
\begin{array}{c}
\smallskip
\disp
\aa_1 t^{-\frac{1}{2}}\tau_3\tau_4+s_1(\tau_1)s_2s_0(\tau_0)-\tau_1 s_1s_2s_0(\tau_0)=0, \\
\smallskip
\disp
\aa_1 t^{-\frac{1}{2}}\tau_0\tau_4+s_1(\tau_1)s_2s_3(\tau_3)-\tau_1 s_1s_2s_3(\tau_3)=0, \\
\smallskip
\disp
\aa_1 t^{-\frac{1}{2}}\tau_0\tau_3+s_1(\tau_1)s_2s_4(\tau_4)-\tau_1 s_1s_2s_4(\tau_4)=0,
\end{array}
\end{equation}
\begin{equation}
\begin{array}{c}
\smallskip
\disp
\aa_3 t^{-\frac{1}{2}}     \tau_0\tau_4-s_3(\tau_3)s_2s_1(\tau_1)+\tau_3 s_3s_2s_1(\tau_1)=0, \\
\smallskip
\disp
\aa_3 t^{-\frac{1}{2}}(1-t)\tau_1\tau_4-s_3(\tau_3)s_2s_0(\tau_0)+\tau_3 s_3s_2s_0(\tau_0)=0, \\
\smallskip
\disp
\aa_3 t^{-\frac{1}{2}}     \tau_0\tau_1-s_3(\tau_3)s_2s_4(\tau_4)+\tau_3 s_3s_2s_4(\tau_4)=0,
\end{array}
\end{equation}
\begin{equation}
\begin{array}{c}
\smallskip
\disp
 \aa_4 t^{-\frac{1}{2}}\tau_0\tau_3-s_4(\tau_4)s_2s_1(\tau_1)+\tau_4 s_4s_2s_1(\tau_1)=0, \\
\smallskip
\disp
-\aa_4 t^{\frac{1}{2}} \tau_1\tau_3-s_4(\tau_4)s_2s_0(\tau_0)+\tau_4 s_4s_2s_0(\tau_0)=0, \\
\smallskip
\disp
-\aa_4 t^{-\frac{1}{2}}\tau_0\tau_1-s_4(\tau_4)s_2s_3(\tau_3)+\tau_4 s_4s_2s_3(\tau_3)=0,
\end{array}
\end{equation}
\begin{equation}
\begin{array}{c}
\smallskip
\disp
\aa_2 t^{-\frac{1}{2}}\tau_3\tau_4-s_1(\tau_1)s_2s_0(\tau_0)+s_0(\tau_0)s_2s_1(\tau_1)=0, \\
\smallskip
\disp
\aa_2 t^{-\frac{1}{2}}\tau_0\tau_4-s_1(\tau_1)s_2s_3(\tau_3)+s_3(\tau_3)s_2s_1(\tau_1)=0, \\
\smallskip
\disp
\aa_2 t^{-\frac{1}{2}}\tau_0\tau_3-s_1(\tau_1)s_2s_4(\tau_4)+s_4(\tau_4)s_2s_1(\tau_1)=0, \\
\smallskip
\disp
\aa_2 t^{-\frac{1}{2}}     \tau_0\tau_1-s_4(\tau_4)s_2s_3(\tau_3)+s_3(\tau_3)s_2s_4(\tau_4)=0, \\
\smallskip
\disp
\aa_2 t^{\frac{1}{2}}      \tau_1\tau_3-s_4(\tau_4)s_2s_0(\tau_0)+s_0(\tau_0)s_2s_4(\tau_4)=0, \\
\disp
\aa_2 t^{-\frac{1}{2}}(t-1)\tau_1\tau_4-s_3(\tau_3)s_2s_0(\tau_0)+s_0(\tau_0)s_2s_3(\tau_3)=0. 
\end{array}   \label{bi:BT:a2}
\end{equation}

\subsection{The $\tau$-functions on the weight lattice of type $D_4$}
Let us define the following translation operators 
\begin{equation}
\begin{array}{c}
 T_{03}=s_3s_0s_2s_4s_1s_2s_6, \quad 
 T_{14}=s_4s_1s_2s_3s_0s_2s_6, \\
 \widehat{T}_{34}=s_3s_2s_0s_1s_2s_3s_5, \quad 
 T_{34}      =s_4s_3s_2s_1s_0s_2s_5, 
\end{array}   \label{def:T}
\end{equation}
which act on parameters $\aa_i$ as 
\begin{equation}
\begin{array}{c}
 T_{03}          (\aa_0,\aa_1,\aa_2,\aa_3,\aa_4)=
                 (\aa_0,\aa_1,\aa_2,\aa_3,\aa_4)+(1,0,-1,1,0), \\
 T_{14}          (\aa_0,\aa_1,\aa_2,\aa_3,\aa_4)=
                 (\aa_0,\aa_1,\aa_2,\aa_3,\aa_4)+(0,1,-1,0,1), \\
 \widehat{T}_{34}(\aa_0,\aa_1,\aa_2,\aa_3,\aa_4)=
                 (\aa_0,\aa_1,\aa_2,\aa_3,\aa_4)+(0,0,0,1,-1), \\
 T_{34}          (\aa_0,\aa_1,\aa_2,\aa_3,\aa_4)=
                 (\aa_0,\aa_1,\aa_2,\aa_3,\aa_4)+(0,0,-1,1,1), 
\end{array}
\end{equation}
and generate the weight lattice of type $D_4$. 
It is possible to derive Toda and Toda-like equations. 

\begin{prop}
We have 
\begin{equation}
\begin{array}{l}
\disp 
T_{03}(\tau_0)T_{03}^{-1}(\tau_0)=t^{-\frac{1}{2}}
\left[(t-1)\frac{d}{dt}(\log \tau_0)'-(\log \tau_0)'
      +\frac{1}{4}(1-\aa_0-\aa_3)^2+\frac{1}{2}\right]\tau_0^2, \\
\disp 
T_{14}(\tau_0)T_{14}^{-1}(\tau_0)=-t^{-\frac{1}{2}}
\left[(t-1)\frac{d}{dt}(\log \tau_0)'-(\log \tau_0)'
      +\frac{1}{4}(\aa_1+\aa_4)^2+\frac{1}{2}\right]\tau_0^2, \\
\disp 
\widehat{T}_{34}(\tau_0)\widehat{T}_{34}^{-1}(\tau_0)=\left(\frac{t-1}{t}\right)^{\frac{1}{2}}
\left[\frac{d}{dt}(\log \tau_0)' +\frac{1}{4}(\aa_3-\aa_4)^2-\frac{1}{2}\right]\tau_0^2, \\
\disp 
T_{34}(\tau_0)T_{34}^{-1}(\tau_0)=\left(\frac{t-1}{t}\right)^{\frac{1}{2}}
\left[\frac{d}{dt}(\log \tau_0)' +\frac{1}{4}(\aa_3+\aa_4)^2-\frac{1}{2}\right]\tau_0^2. 
\end{array}   \label{Toda}
\end{equation}
\end{prop}

\noindent
{\it Proof.}\quad 
Note that 
\begin{equation}
\frac{d}{dt}h_0=-f_2^2f_3f_4+(\aa_3f_4+\aa_4f_3)f_2-\frac{1}{4}(\aa_3+\aa_4)^2+\frac{1}{2}.
\end{equation}
Using (\ref{BT:a}),(\ref{BT:f}),(\ref{BT:tau}),(\ref{s6totau}) and (\ref{def:T}), we have
\begin{equation}
T_{03}(\tau_0)T_{03}^{-1}(\tau_0)=t^{-\frac{1}{2}}
\left[(t-1)\frac{d}{dt}h_0-h_0+\frac{1}{4}(1-\aa_0-\aa_3)^2+\frac{1}{2}\right]\tau_0^2,
\end{equation}
which gives the first equation in (\ref{Toda}). 
The other equations are obtained in similar way. \hfill\qed

\medskip

These translation operators commute each other except for the action on
$\tau$-functions. 
Due to the algebraic relations (\ref{modi:1}) and (\ref{modi:2}), 
the commutation relations on $\tau_0$ are described as 
\begin{equation}
\begin{array}{l}
 T_{03}T_{14}(\tau_0)=-T_{14}T_{03}(\tau_0), \quad 
 T_{03}\widehat{T}_{34}(\tau_0)=i\widehat{T}_{34}T_{03}(\tau_0), \\
 T_{03}T_{34}(\tau_0)=iT_{34}T_{03}(\tau_0), \quad 
 T_{14}\widehat{T}_{34}(\tau_0)=-i\widehat{T}_{34}T_{14}(\tau_0), \\
 T_{14}T_{34}(\tau_0)=-iT_{34}T_{14}(\tau_0), \quad 
 \widehat{T}_{34}T_{34}(\tau_0)=T_{34}\widehat{T}_{34}(\tau_0). 
\end{array}
\end{equation}
We introduce $\tau$-functions on the weight lattice of type $D_4$ as 
\begin{equation}
\tau_{k,l,m,n}=T_{34}^n \widehat{T}_{34}^m T_{14}^l T_{03}^k(\tau_0), \quad 
k,l,m,n \in \BZ. \label{tau}
\end{equation}
In terms of this notation, the $24~\tau$-functions in the bilinear equations
(\ref{bi:norm})-(\ref{bi:BT:a2}) are expressed as follows, 
\begin{equation}
\begin{array}{c}
\medskip
\disp
\tau_{0,0,0,0}=\tau_0, \quad \tau_{1,-1,-1,0}=-[t(t-1)]^{\frac{1}{4}}\tau_1, \quad 
\tau_{1,0,-1,-1}=i t^{\frac{1}{4}}\tau_3, \quad \tau_{1,0,0,-1}=i (t-1)^{\frac{1}{4}}\tau_4, \\
\smallskip
\disp
\tau_{2,0,-1,-1}=-s_0(\tau_0), \quad 
\tau_{1,1,0,-1}=[t(t-1)]^{\frac{1}{4}}s_1(\tau_1), \\
\medskip
\disp
\tau_{1,0,0,0}=i t^{\frac{1}{4}}s_3(\tau_3), \quad 
\tau_{1,0,-1,0}=i(t-1)^{\frac{1}{4}}s_4(\tau_4), \\
\smallskip
\disp
\tau_{1,-1,-1,-1}=s_2s_0(\tau_0), \quad 
\tau_{0,0,0,-1}=-[t(t-1)]^{\frac{1}{4}}s_2s_1(\tau_1), \\
\disp
\tau_{0,-1,0,0}=-i t^{\frac{1}{4}}s_2s_3(\tau_3), \quad 
\tau_{0,-1,-1,0}=i(t-1)^{\frac{1}{4}}s_2s_4(\tau_4), 
\end{array}   \label{init:tau:1}
\end{equation}
\begin{equation}
\begin{array}{c}
\medskip
\disp
\tau_{1,1,0,-2}=-s_1s_2s_0(\tau_0), \quad 
\tau_{1,-1,0,0}=-s_3s_2s_0(\tau_0), \quad 
\tau_{1,-1,-2,0}=s_4s_2s_0(\tau_0), \\
\smallskip
\disp
\tau_{2,0,-1,-2}=[t(t-1)]^{\frac{1}{4}}s_0s_2s_1(\tau_1), \quad 
\tau_{0,0,1,0}=[t(t-1)]^{\frac{1}{4}}s_3s_2s_1(\tau_1), \\
\medskip
\disp
\tau_{0,0,-1,0}=-[t(t-1)]^{\frac{1}{4}}s_4s_2s_1(\tau_1), \\
\medskip
\disp
\tau_{2,-1,-1,-1}=-i t^{\frac{1}{4}}s_0s_2s_3(\tau_3), \quad 
\tau_{0,1,1,-1}=i t^{\frac{1}{4}}s_1s_2s_3(\tau_3), \quad 
\tau_{0,-1,-1,1}=-i t^{\frac{1}{4}}s_4s_2s_3(\tau_3), \\
\smallskip
\disp
\tau_{2,-1,-2,-1}=i(t-1)^{\frac{1}{4}}s_0s_2s_4(\tau_4), \quad 
\tau_{0,1,0,-1}=-i(t-1)^{\frac{1}{4}}s_1s_2s_4(\tau_4), \\
\disp
\tau_{0,-1,0,1}=i(t-1)^{\frac{1}{4}}s_3s_2s_4(\tau_4). 
\end{array}   \label{init:tau:2}
\end{equation}
Toda equations (\ref{Toda}) yield to 
\begin{equation}
\begin{array}{l}
\disp 
\tau_{k+1,l,m,n}\tau_{k-1,l,m,n} \\
\medskip
\disp 
\hskip10pt 
=t^{-\frac{1}{2}}
\left[(t-1)\frac{d}{dt}(\log \tau_{k,l,m,n})'-(\log \tau_{k,l,m,n})'
          +\frac{(1-\aa_0-\aa_3-2k-m-n)^2}{4}+\frac{1}{2}\right]\tau_{k,l,m,n}^2, \\
\disp 
\tau_{k,l+1,m,n}\tau_{k,l-1,m,n} \\
\medskip
\disp 
\hskip10pt 
=-t^{-\frac{1}{2}}
\left[(t-1)\frac{d}{dt}(\log \tau_{k,l,m,n})'-(\log \tau_{k,l,m,n})'
          +\frac{(\aa_1+\aa_4+2l-m+n)^2}{4}+\frac{1}{2}\right]\tau_{k,l,m,n}^2, \\
\medskip
\disp 
\tau_{k,l,m+1,n}\tau_{k,l,m-1,n}
=\left(\frac{t-1}{t}\right)^{\frac{1}{2}}
\left[\frac{d}{dt}(\log \tau_{k,l,m,n})' 
      +\frac{(\aa_3-\aa_4+k-l+2m)^2}{4}-\frac{1}{2}\right]\tau_{k,l,m,n}^2, \\
\disp 
\tau_{k,l,m,n+1}\tau_{k,l,m,n-1}
=\left(\frac{t-1}{t}\right)^{\frac{1}{2}}
\left[\frac{d}{dt}(\log \tau_{k,l,m,n})' 
      +\frac{(\aa_3+\aa_4+k+l+2n)^2}{4}-\frac{1}{2}\right]\tau_{k,l,m,n}^2. 
\end{array}   \label{Toda:tau}
\end{equation}
It is easy to see from (\ref{multi-form}),(\ref{tau}) and (\ref{init:tau:1}) that we have 
\begin{equation}
\begin{array}{l}
\medskip
\disp 
T_{34}^n \widehat{T}_{34}^m T_{14}^l T_{03}^k(f_0)=
t^{\frac{1}{2}}(t-1)^{\frac{1}{2}}\frac{\tau_{k,l,m,n}\tau_{k+2,l,m-1,n-1}}
                       {\tau_{k+1,l-1,m-1,n}\tau_{k+1,l+1,m,n-1}}, \\
\medskip
\disp 
T_{34}^n \widehat{T}_{34}^m T_{14}^l T_{03}^k(f_3)=
(t-1)^{\frac{1}{2}}\frac{\tau_{k+1,l,m-1,n-1}\tau_{k+1,l,m,n}}
                {\tau_{k+1,l-1,m-1,n}\tau_{k+1,l+1,m,n-1}}, \\
\medskip
\disp 
T_{34}^n \widehat{T}_{34}^m T_{14}^l T_{03}^k(f_4)=
t^{\frac{1}{2}}\frac{\tau_{k+1,l,m,n-1}\tau_{k+1,l,m-1,n}}
            {\tau_{k+1,l-1,m-1,n}\tau_{k+1,l+1,m,n-1}}, \\
\disp 
T_{34}^n \widehat{T}_{34}^m T_{14}^l T_{03}^k(f_2)=
-(t-1)^{-\frac{1}{2}}\frac{\tau_{k+1,l-1,m-1,n}\tau_{k+1,l+1,m,n-1}\tau_{k,l,m,n-1}}
                  {\tau_{k,l,m,n}\tau_{k+1,l,m-1,n-1}\tau_{k+1,l,m,n-1}}. 
\end{array}   \label{f:BT}
\end{equation}

\section{Construction of a family of algebraic solutions \label{constr}}
It is known that one can get an algebraic solution of Painlev\'e
equations by considering the fixed points with respect to the B\"acklund
transformations corresponding to Dynkin automorphisms~\cite{Um,NY:P4}. 
Iteration of B\"acklund transformations to the seed solution gives a
family of algebraic solutions, 
which are expressed by the ratio of some characteristic polynomials,
such as Yablonskii-Vorob'ev, Okamoto and Umemura polynomials. 
These polynomials are defined as the non-trivial factors of
$\tau$-functions and generated by Toda type recursion relations. 

In this section, we construct a family of algebraic solutions to the
symmetric form of P$_{\rm VI}$ by following the above recipe.

\subsection{A seed solution}
Consider the Dynkin diagram automorphism $s_6$ to get a seed solution. 
By (\ref{s6}), the fixed solution is derived from 
\begin{equation}
\aa_0=\aa_3, \quad \aa_1=\aa_4, \quad 
f_4=\frac{t}{f_4}, \quad f_2=-\frac{f_4(f_4f_2+\aa_2)}{t}. 
\end{equation}
Then, we obtain 
\begin{equation}
\begin{array}{c}
\smallskip
\disp 
(\aa_0,\aa_1,\aa_2,\aa_3,\aa_4)=\left(a,b,\frac{1}{2}-a-b,a,b \right), \\
\disp 
f_0=x-x^2, \quad f_3=x-1, \quad f_4=x, \quad 
f_2=\frac{1}{2}\left(a+b-\frac{1}{2}\right)x^{-1}, \quad 
x^2=t, 
\end{array}   \label{seed}
\end{equation}
as a seed solution, which is equivalent to the following algebraic
solution of S$_{\rm VI}$, 
\begin{equation}
q=x, \quad p=\frac{1}{2}\left(a+b-\frac{1}{2}\right)x^{-1}, 
\end{equation}
for the parameters 
\begin{equation}
\kappa_{\infty}=b, \quad \kappa_0=b, \quad \kappa_1=a, \quad \theta=a. \label{plane}
\end{equation}

\begin{rem}
Equation (\ref{plane}) gives a plane in the parameter space. 
One can choose the other diagram automorphism, e.g., $s_5$, 
to get a seed solution.
Such a solution exists on the other plane and can be transformed to
 (\ref{seed}) by some B\"acklund transformations. 
The seed solution (\ref{seed}) is the simplest one. 
\end{rem}

Under the specialization of (\ref{seed}), the Hamiltonians $h_i$ and
$\tau$-functions $\tau_i$ are calculated as 
\begin{eqnarray}
&&h_0= \frac{7}{16}x^2+\frac{1}{8}(2a+2b-1)(2a-2b-1)x+\frac{1}{16}(8a^2-8a+3+8b^2), \cr
&&h_1=-\frac{1}{16}x^2+\frac{1}{8}(2a+2b-1)(2a-2b+1)x+\frac{1}{16}(8a^2+8b^2-8b+7), \cr
&&h_2=-\frac{1}{8} x^2+\frac{1}{4}(4a^2-4b^2-1)     x+\frac{1}{8} (8a^2+8b^2+7),    \\
&&h_3= \frac{3}{16}x^2+\frac{1}{8}(2a+2b-1)(2a-2b-1)x+\frac{1}{16}(8a^2-8a+7+8b^2), \cr
&&h_4= \frac{3}{16}x^2+\frac{1}{8}(2a+2b-1)(2a-2b+1)x+\frac{1}{16}(8a^2+8b^2-8b+3), 
\nonumber 
\end{eqnarray}
and 
\begin{equation}
\begin{array}{l}
\medskip
\disp 
\tau_0=(x-1)^{a^2-a+\frac{3}{4}}x^{-a^2+a-b^2-\frac{3}{8}}(x+1)^{b^2+\frac{1}{2}},    \\
\medskip
\disp 
\tau_1=(x-1)^{a^2+\frac{1}{4}}  x^{-a^2-b^2+b-\frac{7}{8}}(x+1)^{b^2-b+\frac{1}{2}}, \\
\medskip
\disp 
\tau_2=(x-1)^{2a^2+\frac{1}{2}} x^{-2a^2-2b^2-\frac{7}{4}}(x+1)^{2b^2+1},        \\
\medskip
\disp 
\tau_3=(x-1)^{a^2-a+\frac{3}{4}}x^{-a^2+a-b^2-\frac{7}{8}}(x+1)^{b^2+\frac{1}{2}},    \\
\disp 
\tau_4=(x-1)^{a^2+\frac{1}{4}}  x^{-a^2-b^2+b-\frac{3}{8}}(x+1)^{b^2-b+\frac{1}{2}}, 
\end{array}
\end{equation}
up to the multiplication by some constants, respectively. 

Using the multiplicative formulas (\ref{multi-form}) and the bilinear
equations (\ref{bi:norm})-(\ref{bi:BT:a2}), 
we get the $24~\tau$-functions in (\ref{init:tau:1}) and (\ref{init:tau:2}). 
These are expressed in the form of 
\begin{equation}
\begin{array}{c}
\medskip
\disp 
\tau_{k,l,m,n}=
\sigma_{k,l,m,n}
(x-1)^{\tilde{a}^2+\frac{1}{2}}
    x^{-\tilde{a}^2-\tilde{b}^2-\frac{1}{2}m(m+1)-\frac{1}{2}n(n+1)-\frac{1}{8}}
(x+1)^{\tilde{b}^2+\frac{1}{2}}, \\
\disp 
\tilde{a}=a+k+\frac{m+n-1}{2}, \quad \tilde{b}=b+l+\frac{-m+n}{2},
\end{array}   \label{tau-sig}
\end{equation}
where $\sigma_{k,l,m,n}$ are given as follows, 
\begin{equation}
\sigma_{0,0,0,0}=1, \quad \sigma_{1,0,0,0}=i, \quad 
\sigma_{0,-1,0,0}=- \frac{i}{2}\left(\frac{1}{2}-a-b\right), \quad 
\sigma_{1,-1,0,0}=- \frac{1}{2}\left(\frac{1}{2}+a-b\right),  \label{init-data}
\end{equation}
\begin{equation}
\sigma_{0,0,-1,0}= \frac{1}{2}\left(\frac{1}{2}-a+b\right), \quad 
\sigma_{0,-1,-1,0}= \frac{i}{2}\left(\frac{1}{2}-a-b\right), \quad 
\sigma_{1,0,-1,0}=i, \quad \sigma_{1,-1,-1,0}=-1, 
\end{equation}
\begin{equation}
\sigma_{0,0,0,-1}= \frac{1}{2}\left(\frac{1}{2}-a-b\right), \quad 
\sigma_{0,1,0,-1}=- \frac{i}{2}\left(\frac{1}{2}-a+b\right), \quad 
\sigma_{1,0,0,-1}=i, \quad \sigma_{1,1,0,-1}=1, 
\end{equation}
\begin{equation}
\sigma_{1,0,-1,-1}=i, \quad \sigma_{2,0,-1,-1}=1, \quad 
\sigma_{1,-1,-1,-1}= \frac{1}{2}\left(\frac{1}{2}-a-b\right), \quad 
\sigma_{2,-1,-1,-1}= \frac{i}{2}\left(\frac{1}{2}+a-b\right), 
\end{equation}
\begin{equation}
\begin{array}{c}
\medskip
\disp
\sigma_{0,-1,0,1}=
\frac{i}{2}
\left[\left(\frac{1}{2}-a-b\right)x
     -\left(\frac{1}{2}+a-b\right)\right],\\
\medskip
\disp
\sigma_{0,0,1,0}=
-\frac{1}{2}
\left[\left(\frac{1}{2}+a-b\right)x
     -\left(\frac{1}{2}-a-b\right)\right], 
\end{array}
\end{equation}
\begin{equation}
\begin{array}{c}
\medskip
\disp
\sigma_{0,1,1,-1}= 
\frac{i}{2}
\left[\left(\frac{1}{2}-a+b\right)x
     +\left(\frac{1}{2}-a-b\right)\right],\\
\medskip
\disp
\sigma_{0,-1,-1,1}= 
-\frac{i}{2}
\left[\left(\frac{1}{2}-a-b\right)x
     +\left(\frac{1}{2}-a+b\right)\right], 
\end{array}
\end{equation}
\begin{equation}
\begin{array}{c}
\medskip
\disp
\sigma_{1,1,0,-2}= 
-\frac{1}{2}
\left[\left(\frac{1}{2}-a-b\right)x
     +\left(\frac{1}{2}-a+b\right)\right],\\
\medskip
\disp
\sigma_{1,-1,-2,0}= 
\frac{1}{2}
\left[\left(\frac{1}{2}-a+b\right)x
     +\left(\frac{1}{2}-a-b\right)\right], 
\end{array}
\end{equation}
\begin{equation}
\begin{array}{c}
\medskip
\disp
\sigma_{2,-1,-2,-1}= 
-\frac{i}{2}
\left[\left(\frac{1}{2}+a-b\right)x
     -\left(\frac{1}{2}-a-b\right)\right],\\
\disp
\sigma_{2,0,-1,-2}= 
\frac{1}{2}
\left[\left(\frac{1}{2}-a-b\right)x
     -\left(\frac{1}{2}+a-b\right)\right]. 
\end{array}   \label{init-data'}
\end{equation}

\subsection{Application of B\"acklund transformations}
Assume that $\tau_{k,l,m,n}$ are expressed as (\ref{tau-sig}) for any
$k,l,m,n \in \BZ$. 
Substituting $\aa_0=\aa_3=a,~\aa_1=\aa_4=b$ and (\ref{tau-sig}) into
(\ref{Toda:tau}), 
we obtain the Toda equations for $\sigma_{k,l,m,n}$. 
The first two equations yield to 
\begin{equation}
\begin{array}{c}
\medskip
\disp 
4\sigma_{k+1,l,m,n}\sigma_{k-1,l,m,n}=
\left[(x+1)^2{\cal D}^2-(\tilde{a}+\tilde{b})(\tilde{a}-\tilde{b})\right]
\sigma_{k,l,m,n}\cdot \sigma_{k,l,m,n}, \\
\disp 
4\sigma_{k,l+1,m,n}\sigma_{k,l-1,m,n}=
-\left[(x-1)^2{\cal D}^2-(\tilde{a}+\tilde{b})(\tilde{a}-\tilde{b})\right]
\sigma_{k,l,m,n}\cdot \sigma_{k,l,m,n}, 
\end{array}   \label{Toda:sig'}
\end{equation}
where we denote as 
\begin{equation}
{\cal D}^2 \sigma \cdot \sigma=
x(\ddot{\sigma}\sigma-\dot{\sigma}^2)+\dot{\sigma}\sigma, \quad 
\dot{\sigma}=\frac{d \sigma}{dx}. 
\end{equation}
The others are reduced to 
\begin{equation}
\begin{array}{l}
\medskip
\disp 
4\sigma_{k,l,m+1,n}\sigma_{k,l,m-1,n}=
(x-1)x(x+1)(\ddot{\sigma}_{k,l,m,n}\sigma_{k,l,m,n}
-\dot{\sigma}_{k,l,m,n}^2)+(3x^2-1)\dot{\sigma}_{k,l,m,n}\sigma_{k,l,m,n} \\
\medskip
\disp 
\hskip100pt 
+\left\{\left[\left(\hat{a}-\hat{b}+m-\frac{1}{2}\right)
              \left(\hat{a}-\hat{b}+3m+\frac{1}{2}\right)-n(n+1)\right]x \right. \\
\medskip
\disp 
\hskip190pt \left. 
+\left(\hat{a}-\hat{b}+m-\frac{1}{2}\right)
 \left(\hat{a}+\hat{b}+n-\frac{1}{2}\right)\right\}\sigma_{k,l,m,n}^2, \\
\medskip
\disp 
4\sigma_{k,l,m,n+1}\sigma_{k,l,m,n-1}=
(x-1)x(x+1)(\ddot{\sigma}_{k,l,m,n}\sigma_{k,l,m,n}
-\dot{\sigma}_{k,l,m,n}^2)+(3x^2-1)\dot{\sigma}_{k,l,m,n}\sigma_{k,l,m,n} \\
\medskip
\disp 
\hskip100pt 
+\left\{\left[\left(\hat{a}+\hat{b}+n-\frac{1}{2}\right)
              \left(\hat{a}+\hat{b}+3n+\frac{1}{2}\right)-m(m+1)\right]x \right. \\
\disp 
\hskip190pt \left. 
+\left(\hat{a}+\hat{b}+n-\frac{1}{2}\right)
 \left(\hat{a}-\hat{b}+m-\frac{1}{2}\right)\right\}\sigma_{k,l,m,n}^2, 
\end{array}   \label{Toda:sig}
\end{equation}
with
\begin{equation}
\hat{a}=a+k, \quad \hat{b}=b+l. 
\end{equation}
Toda equations (\ref{Toda:sig'}) and (\ref{Toda:sig}) with the initial
data (\ref{init-data})-(\ref{init-data'}) generate
$\sigma_{k,l,m,n}=\sigma_{k,l,m,n}(x;a,b)$ for $k,l,m,n \in \BZ$. 
From (\ref{f:BT}) and (\ref{tau-sig}), we see that the ratio of
$\sigma_{k,l,m,n}$ gives a family of algebraic solutions to the
symmetric form of P$_{\rm VI}$. 

Noticing that we have 
\begin{equation}
T_{14}^lT_{03}^k(\aa_0,\aa_1,\aa_2,\aa_3,\aa_4)=
\left(\hat{a},\hat{b},\frac{1}{2}-\hat{a}-\hat{b},\hat{a},\hat{b}\right), 
\end{equation}
under the specialization (\ref{seed}), 
we see that the action of $T_{03}$ and $T_{14}$ on the parameter space
is absorbed by shift of the parameters $a$ and $b$, respectively. 
This suggests that we do not need to consider the translations $T_{03}$
and $T_{14}$ in order to get a family of algebraic solutions of 
P$_{\rm VI}$. 

To verify this, we put 
\begin{equation}
\sigma_{k,l,m,n}=\omega_{k,l,m,n}V_{k,l,m,n}, \quad 
\omega_{k,l,m,n}=\omega_{k,l,m,n}(a,b),  \quad k,l,m,n \in \BZ. 
\label{sig-V}
\end{equation}
The constants $\omega_{k,l,m,n}$ are determined by recurrence relations
as follows. 
With respect to the indices $k$ and $l$, $\omega_{k,l,i,j}$ with
$(i,j)=(-1,-1),(-1,0),(0,-1),(0,0)$ are subject to 
\begin{equation}
\begin{array}{c}
\medskip
\disp 
4\omega_{k+1,l,i,j}\omega_{k-1,l,i,j}=
-\left(\hat{a}-\hat{b}+i-\frac{1}{2}\right)
 \left(\hat{a}+\hat{b}+j-\frac{1}{2}\right)\omega_{k,l,i,j}^2, \\
\disp 
4\omega_{k,l+1,i,j}\omega_{k,l-1,i,j}=
 \left(\hat{a}-\hat{b}+i-\frac{1}{2}\right)
 \left(\hat{a}+\hat{b}+j-\frac{1}{2}\right)\omega_{k,l,i,j}^2.
\end{array}
\end{equation}
The initial conditions are given by 
\begin{equation}
\omega_{0,0,0,0}=1, \quad \omega_{1,0,0,0}=i \quad 
\omega_{0,-1,0,0}=-\frac{i}{2}\left(\frac{1}{2}-a-b\right), \quad 
\omega_{1,-1,0,0}=-\frac{1}{2}\left(\frac{1}{2}+a-b\right), 
\end{equation}
\begin{equation}
\omega_{0,0,-1,0}=\frac{1}{2}\left(\frac{1}{2}-a+b\right), \quad 
\omega_{0,-1,-1,0}=\frac{i}{2}\left(\frac{1}{2}-a-b\right), \quad 
\omega_{1,0,-1,0}=i, \quad \omega_{1,-1,-1,0}=-1, 
\end{equation}
\begin{equation}
\omega_{1,0,0,-1}=i, \quad \omega_{1,1,0,-1}=1 \quad 
\omega_{0,0,0,-1}=\frac{1}{2}\left(\frac{1}{2}-a-b\right), \quad 
\omega_{0,1,0,-1}=-\frac{i}{2}\left(\frac{1}{2}-a+b\right), 
\end{equation}
\begin{equation}
\omega_{1,0,-1,-1}=i, \quad \omega_{2,0,-1,-1}=1 \quad 
\omega_{1,-1,-1,-1}=\frac{1}{2}\left(\frac{1}{2}-a-b\right), \quad 
\omega_{2,-1,-1,-1}=\frac{i}{2}\left(\frac{1}{2}+a-b\right). 
\end{equation}
Note that these imply 
\begin{equation}
V_{k,l,-1,-1}=V_{k,l,-1,0}=V_{k,l,0,-1}=V_{k,l,0,0}=1, \quad k,l \in \BZ. \label{init'}
\end{equation}
For the indices $m$ and $n$, we set 
\begin{equation}
\begin{array}{c}
\medskip
\disp 
8\omega_{k,l,m+1,n}\omega_{k,l,m-1,n}=
\left(\hat{a}-\hat{b}+m-\frac{1}{2}\right)\omega_{k,l,m,n}^2, \\
\disp 
8\omega_{k,l,m,n+1}\omega_{k,l,m,n-1}=
\left(\hat{a}+\hat{b}+n-\frac{1}{2}\right)\omega_{k,l,m,n}^2. 
\end{array}
\end{equation}
Thus, $\omega_{k,l,m,n}$ are determined for any $k,l,m,n\in\BZ$. 
As a result, the functions $V_{k,l,m,n}=V_{k,l,m,n}(x;a,b)$ introduced
in (\ref{sig-V}) have a symmetry described in the following lemma. 

\begin{lem}
We have 
\begin{equation}
V_{k,l,m,n}(x;a,b)=V_{0,0,m,n}(x;a+k,b+l). \label{V-shift}
\end{equation}
\end{lem}

\noindent
{\it Proof.}\quad 
Toda equations (\ref{Toda:sig}) are reduced to 
\begin{eqnarray}
&&\hskip-20pt
\frac{1}{2}\left(\hat{a}-\hat{b}+m-\frac{1}{2}\right)V_{k,l,m+1,n}V_{k,l,m-1,n} \cr
&&=(x-1)x(x+1)(\ddot{V}_{k,l,m,n}V_{k,l,m,n}-\dot{V}_{k,l,m,n}^2)
  +(3x^2-1)\dot{V}_{k,l,m,n}V_{k,l,m,n} \cr
&&\hskip10pt 
+\left\{\left(\hat{a}-\hat{b}+m-\frac{1}{2}\right)
  \left[\left(\hat{a}-\hat{b}+3m+\frac{1}{2}\right)x
       +\left(\hat{a}+\hat{b}+n-\frac{1}{2}\right)\right]
  -n(n+1)x\right\}V_{k,l,m,n}^2, \cr 
&& \\
&&\hskip-20pt
\frac{1}{2}\left(\hat{a}+\hat{b}+n-\frac{1}{2}\right)V_{k,l,m,n+1}V_{k,l,m,n-1} \cr
&&=(x-1)x(x+1)(\ddot{V}_{k,l,m,n}V_{k,l,m,n}-\dot{V}_{k,l,m,n}^2)
  +(3x^2-1)\dot{V}_{k,l,m,n}V_{k,l,m,n} \cr
&&\hskip10pt
+\left\{\left(\hat{a}+\hat{b}+n-\frac{1}{2}\right)
  \left[\left(\hat{a}+\hat{b}+3n+\frac{1}{2}\right)x
       +\left(\hat{a}-\hat{b}+m-\frac{1}{2}\right)\right]
  -m(m+1)x\right\}V_{k,l,m,n}^2. \nonumber 
\end{eqnarray}
Then, we see that $V_{k,l,m,n}(x;a,b)$ satisfy the same Toda equations
as $V_{0,0,m,n}(x;a+k,b+l)$. 
Since the initial conditions are given by (\ref{init'}), 
we obtain (\ref{V-shift}). \hfill\qed

\medskip

On the other hand, we have from (\ref{f:BT}),(\ref{tau-sig}) and (\ref{sig-V}) 
\begin{eqnarray}
\hskip-20pt 
T_{34}^n \widehat{T}_{34}^m T_{14}^l T_{03}^k(f_0)
&=&x(x-1)\frac{\sigma_{k,l,m,n}\sigma_{k+2,l,m-1,n-1}}
              {\sigma_{k+1,l-1,m-1,n}\sigma_{k+1,l+1,m,n-1}} \cr
&=&x(x-1)\frac{\omega_{k,l,m,n}\omega_{k+2,l,m-1,n-1}}
              {\omega_{k+1,l-1,m-1,n}\omega_{k+1,l+1,m,n-1}}
         \frac{V_{k,l,m,n}V_{k+2,l,m-1,n-1}}
              {V_{k+1,l-1,m-1,n}V_{k+1,l+1,m,n-1}}. 
\end{eqnarray}
The ratio of $\omega$'s is calculated as 
\begin{equation}
\frac{\omega_{k,l,m,n}\omega_{k+2,l,m-1,n-1}}
     {\omega_{k+1,l-1,m-1,n}\omega_{k+1,l+1,m,n-1}}=-1. 
\end{equation}
Similarly, for $f_3,f_4$ and $f_2$, we get 
\begin{equation}
\begin{array}{c}
\medskip
\disp 
\frac{\omega_{k+1,l,m-1,n-1}\omega_{k+1,l,m,n}}
     {\omega_{k+1,l-1,m-1,n}\omega_{k+1,l+1,m,n-1}}=1, \quad 
\frac{\omega_{k+1,l,m,n-1}\omega_{k+1,l,m-1,n}}
     {\omega_{k+1,l-1,m-1,n}\omega_{k+1,l+1,m,n-1}}=1, \\ 
\disp 
\frac{\omega_{k+1,l-1,m-1,n}\omega_{k+1,l+1,m,n-1}\omega_{k,l,m,n-1}}
     {\omega_{k,l,m,n}\omega_{k+1,l,m-1,n-1}\omega_{k+1,l,m,n-1}}=
-\frac{1}{2}\left(\hat{a}+\hat{b}+n-\frac{1}{2}\right). 
\end{array}
\end{equation}
The above discussion means that we can put $k=l=0$ without loss of
generality for our purpose. 
Hence, we denote $V_{0,0,m,n}=V_{m,n}$. 

Moreover, it is observed that $V_{m,n}=V_{m,n}(x;a,b)~(m,n\in\BZ)$ are
polynomials in $a,b$ and $x$ with coefficients in $\BZ$. 
We will show this in the next section by presenting the explicit
expressions. 

\begin{prop}
Let $V_{m,n}=V_{m,n}(x;a,b)~(m,n \in \BZ)$ be polynomials generated by Toda equations, 
\begin{eqnarray}
&&\hskip-20pt
\frac{1}{2}\left(a-b+m-\frac{1}{2}\right)V_{m+1,n}V_{m-1,n} \cr
&&=(x-1)x(x+1)(\ddot{V}_{m,n}V_{m,n}-\dot{V}_{m,n}^2)
  +(3x^2-1)\dot{V}_{m,n}V_{m,n} \cr
&&\hskip10pt
+\left\{\left(a-b+m-\frac{1}{2}\right)
  \left[\left(a-b+3m+\frac{1}{2}\right)x
        +\left(a+b+n-\frac{1}{2}\right)\right]
  -n(n+1)x\right\}V_{m,n}^2, \cr
&& \label{Toda:V} \\
&&\hskip-20pt
\frac{1}{2}\left(a+b+n-\frac{1}{2}\right)V_{m,n+1}V_{m,n-1} \cr
&&=(x-1)x(x+1)(\ddot{V}_{m,n}V_{m,n}-\dot{V}_{m,n}^2)
  +(3x^2-1)\dot{V}_{m,n}V_{m,n} \cr
&&\hskip10pt
+\left\{\left(a+b+n-\frac{1}{2}\right)
  \left[\left(a+b+3n+\frac{1}{2}\right)x
        +\left(a-b+m-\frac{1}{2}\right)\right]
  -m(m+1)x\right\}V_{m,n}^2, \nonumber 
\end{eqnarray}
with the initial conditions, 
\begin{equation}
V_{-1,-1}=V_{-1,0}=V_{0,-1}=V_{0,0}=1. \label{init}
\end{equation}
Then, 
\begin{equation}
\begin{array}{l}
\medskip
\disp 
f_0=x(1-x)\frac{V_{m,n}(x;a,b)V_{m-1,n-1}(x;a+2,b)}
               {V_{m-1,n}(x;a+1,b-1)V_{m,n-1}(x;a+1,b+1)}, \\
\medskip
\disp 
f_3=(x-1)\frac{V_{m-1,n-1}(x;a+1,b)V_{m,n}(x;a+1,b)}
              {V_{m-1,n}(x;a+1,b-1)V_{m,n-1}(x;a+1,b+1)}, \\
\medskip
\disp 
f_4=x\frac{V_{m,n-1}(x;a+1,b)V_{m-1,n}(x;a+1,b)}
          {V_{m-1,n}(x;a+1,b-1)V_{m,n-1}(x;a+1,b+1)}, \\ 
\disp 
f_2=\frac{1}{2}\left(a+b+n-\frac{1}{2}\right)x^{-1}
    \frac{V_{m-1,n}(x;a+1,b-1)V_{m,n-1}(x;a+1,b+1)V_{m,n-1}(x;a,b)}
         {V_{m,n}(x;a,b)V_{m-1,n-1}(x;a+1,b)V_{m,n-1}(x;a+1,b)}, 
\end{array}   \label{f-V}
\end{equation}
satisfy the symmetric form of P$_{\rm VI}$ for the parameters 
\begin{equation}
(\aa_0,\aa_1,\aa_2,\aa_3,\aa_4)=\left(a,b,\frac{1}{2}-a-b-n,a+m+n,b-m+n\right). 
\label{para:sym}
\end{equation}
\end{prop}
Furthermore, the bilinear relations for $V_{m,n}$ are derived from
(\ref{bi:norm})-(\ref{bi:BT:a2}). 
\begin{prop}
The polynomials $V_{m,n}(x;a,b)$ satisfy the following bilinear
 relations, 
\begin{equation}
\begin{array}{l}
\medskip
\disp 
V_{m-1,n}^{(1,-1)}V_{m,n-1}^{(1,1)}
+(x-1)V_{m-1,n-1}^{(1,0)}V_{m,n}^{(1,0)}-xV_{m,n-1}^{(1,0)}V_{m-1,n}^{(1,0)}=0, \\
\medskip
\disp 
(x-1)V_{m,n}^{(0,0)}V_{m-1,n-1}^{(2,0)}
-xV_{m-1,n}^{(1,-1)}V_{m,n-1}^{(1,1)}+V_{m,n-1}^{(1,0)}V_{m-1,n}^{(1,0)}=0, \\
\medskip
\disp 
V_{m-1,n}^{(1,-1)}V_{m,n-1}^{(0,0)}
-(x+1)V_{m-1,n-1}^{(1,0)}V_{m,n}^{(0,-1)}+xV_{m,n-1}^{(1,0)}V_{m-1,n}^{(0,-1)}=0, \\
\medskip
\disp 
(x+1)V_{m,n}^{(0,0)}V_{m-1,n-1}^{(1,-1)}
-xV_{m-1,n}^{(1,-1)}V_{m,n-1}^{(0,0)}-V_{m,n-1}^{(1,0)}V_{m-1,n}^{(0,-1)}=0, 
\end{array}   \label{bi:norm'}
\end{equation}
\begin{equation}
\begin{array}{l}
\medskip
\disp 
4aV_{m-1,n-1}^{(1,0)}V_{m,n-1}^{(1,0)}
+2\nu_n(x-1)V_{m-1,n-1}^{(2,0)}V_{m,n-1}^{(0,0)}
-V_{m,n}^{(0,0)}V_{m-1,n-2}^{(2,0)}=0, \\
\medskip
\disp 
2aV_{m-1,n}^{(1,-1)}V_{m,n-1}^{(1,0)}
-\nu_n V_{m-1,n-1}^{(2,0)}V_{m,n}^{(0,-1)}
-\mu_{m+1}V_{m,n}^{(0,0)}V_{m-1,n-1}^{(2,-1)}=0, \\
\medskip
\disp 
4axV_{m-1,n}^{(1,-1)}V_{m-1,n-1}^{(1,0)}
-2\nu_n(x-1)V_{m-1,n-1}^{(2,0)}V_{m-1,n}^{(0,-1)}
-V_{m,n}^{(0,0)}V_{m-2,n-1}^{(2,-1)}=0, 
\end{array}
\end{equation}
\begin{equation}
\begin{array}{l}
\medskip
\disp 
4bV_{m-1,n-1}^{(1,0)}V_{m,n-1}^{(1,0)}
-2\nu_n(x+1)V_{m,n-1}^{(1,1)}V_{m-1,n-1}^{(1,-1)}
+V_{m-1,n}^{(1,-1)}V_{m,n-2}^{(1,1)}=0, \\
\medskip
\disp 
4bxV_{m,n}^{(0,0)}V_{m,n-1}^{(1,0)}
-2\nu_n(x+1)V_{m,n-1}^{(1,1)}V_{m,n}^{(0,-1)}
+V_{m-1,n}^{(1,-1)}V_{m+1,n-1}^{(0,1)}=0, \\
\medskip
\disp 
2bV_{m,n}^{(0,0)}V_{m-1,n-1}^{(1,0)}
-\nu_n V_{m,n-1}^{(1,1)}V_{m-1,n}^{(0,-1)}
+\mu_m V_{m-1,n}^{(1,-1)}V_{m,n-1}^{(0,1)}=0, 
\end{array}
\end{equation}
\begin{equation}
\begin{array}{l}
\medskip
\disp 
4(a+m+n)xV_{m,n}^{(0,0)}V_{m,n-1}^{(1,0)}
-2\nu_n(x-1)V_{m,n}^{(1,0)}V_{m,n-1}^{(0,0)}
-V_{m-1,n-1}^{(1,0)}V_{m+1,n}^{(0,0)}=0, \\
\medskip
\disp 
2(a+m+n)V_{m-1,n}^{(1,-1)}V_{m,n-1}^{(1,0)}
-\nu_n V_{m,n}^{(1,0)}V_{m-1,n-1}^{(1,-1)}
-\mu_{m+1}V_{m-1,n-1}^{(1,0)}V_{m,n}^{(1,-1)}=0, \\
\medskip
\disp 
4(a+m+n)V_{m,n}^{(0,0)}V_{m-1,n}^{(1,-1)}
+2\nu_n(x-1)V_{m,n}^{(1,0)}V_{m-1,n}^{(0,-1)}
-V_{m-1,n-1}^{(1,0)}V_{m,n+1}^{(0,-1)}=0, 
\end{array}
\end{equation}
\begin{equation}
\begin{array}{l}
\medskip
\disp 
2(b-m+n)V_{m,n}^{(0,0)}V_{m-1,n-1}^{(1,0)}
-\nu_n V_{m-1,n}^{(1,0)}V_{m,n-1}^{(0,0)}
+\mu_m V_{m,n-1}^{(1,0)}V_{m-1,n}^{(0,0)}=0, \\
\medskip
\disp 
4(b-m+n)xV_{m-1,n}^{(1,-1)}V_{m-1,n-1}^{(1,0)}
-2\nu_n(x+1)V_{m-1,n}^{(1,0)}V_{m-1,n-1}^{(1,-1)}
+V_{m,n-1}^{(1,0)}V_{m-2,n}^{(1,-1)}=0, \\
\medskip
\disp 
4(b-m+n)V_{m,n}^{(0,0)}V_{m-1,n}^{(1,-1)}
-2\nu_n(x+1)V_{m-1,n}^{(1,0)}V_{m,n}^{(0,-1)}
+V_{m,n-1}^{(1,0)}V_{m-1,n+1}^{(0,-1)}=0, 
\end{array}
\end{equation}
\begin{equation}
\begin{array}{l}
\medskip
\disp 
2V_{m-1,n-1}^{(1,0)}V_{m,n-1}^{(1,0)}
-(x+1)V_{m,n-1}^{(1,1)}V_{m-1,n-1}^{(1,-1)}
+(x-1)V_{m-1,n-1}^{(2,0)}V_{m,n-1}^{(0,0)}=0, \\
\medskip
\disp 
2xV_{m,n}^{(0,0)}V_{m,n-1}^{(1,0)}
-(x+1)V_{m,n-1}^{(1,1)}V_{m,n}^{(0,-1)}
-(x-1)V_{m,n}^{(1,0)}V_{m,n-1}^{(0,0)}=0, \\
\medskip
\disp 
2V_{m,n}^{(0,0)}V_{m-1,n-1}^{(1,0)}
-V_{m,n-1}^{(1,1)}V_{m-1,n}^{(0,-1)}
-V_{m-1,n}^{(1,0)}V_{m,n-1}^{(0,0)}=0, \\
\medskip
\disp 
2V_{m,n}^{(0,0)}V_{m-1,n}^{(1,-1)}
-(x+1)V_{m-1,n}^{(1,0)}V_{m,n}^{(0,-1)}
+(x-1)V_{m,n}^{(1,0)}V_{m-1,n}^{(0,-1)}=0, \\
\medskip
\disp 
2xV_{m-1,n}^{(1,-1)}V_{m-1,n-1}^{(1,0)}
-(x+1)V_{m-1,n}^{(1,0)}V_{m-1,n-1}^{(1,-1)}
-(x-1)V_{m-1,n-1}^{(2,0)}V_{m-1,n}^{(0,-1)}=0, \\
\disp 
2V_{m-1,n}^{(1,-1)}V_{m,n-1}^{(1,0)}
-V_{m,n}^{(1,0)}V_{m-1,n-1}^{(1,-1)}
-V_{m-1,n-1}^{(2,0)}V_{m,n}^{(0,-1)}=0, 
\end{array}   \label{bi:BT:a2'}
\end{equation}
where we denote as 
\begin{equation}
\mu_m=a-b+m-\frac{1}{2}, \quad \nu_n=a+b+n-\frac{1}{2}, 
\end{equation}
\begin{equation}
V_{m,n}^{(k,l)}=V_{m,n}(x;a+k,b+l). 
\end{equation}
\end{prop}

Conversely, by solving the bilinear relations
(\ref{bi:norm'})-(\ref{bi:BT:a2'}) with the initial conditions
(\ref{init}), 
one can get the family of algebraic solutions (\ref{f-V}) with
 (\ref{para:sym}). 
Even though these bilinear relations are overdetermined systems, 
the consistency is guaranteed by construction. 
In order to show Theorem \ref{detV}, 
we will prove not Toda equations (\ref{Toda:V}) but the bilinear
relations (\ref{bi:norm'})-(\ref{bi:BT:a2'}).

\section{Proof of the determinant formula \label{proof}}
In this section, we give a proof of Theorem \ref{detV}. 
\begin{prop}\label{VtoS:V}
We have 
\begin{equation}
V_{m,n}(x;a,b)=(-2x)^{m(m+1)/2}(-2)^{n(n+1)/2}\xi_m \xi_n S_{m,n}(x;a,b), 
\quad m,n \in \BZ, \label{V-S:V}
\end{equation}
where $S_{m,n}=S_{m,n}(x;a,b)$ is defined in Theorem \ref{detV} and
 $\xi_n$ is the factor determined by 
\begin{equation}
\xi_{n+1}\xi_{n-1}=(2n+1)\xi_n^2, \quad \xi_{-1}=\xi_0=1. \label{xi}
\end{equation}
\end{prop}
From this Proposition, 
it is easy to verify that $V_{m,n}(x;a,b)$ are indeed polynomials in
$a,b$ and $x$ with coefficients in $\BZ$. 

Substituting (\ref{V-S:V}) into (\ref{f-V}), 
we find that Theorem \ref{detV} is a direct consequence of Proposition
\ref{VtoS:V}. 
Taking (\ref{S-R:V}),(\ref{abcd:V}) and (\ref{V-S:V}) into account, 
we obtain the bilinear relations for $R_{m,n}$. 

\begin{prop}\label{bi:rel:V}
Let $R_{m,n}$ be a family of polynomials given in Definition \ref{def:V}. 
Then, we have 
\begin{equation}
\begin{array}{l}
\medskip
\disp 
R_{m-1,n}^{(0,-1)}R_{m,n-1}^{(1,1)}
+(x-1)R_{m-1,n-1}^{(0,0)}R_{m,n}^{(1,0)}
-xR_{m,n-1}^{(0,-1)}R_{m-1,n}^{(1,1)}=0, \\
\medskip
\disp 
(x-1)R_{m,n}^{(0,0)}R_{m-1,n-1}^{(1,0)}
-xR_{m-1,n}^{(0,-1)}R_{m,n-1}^{(1,1)}
+R_{m,n-1}^{(0,-1)}R_{m-1,n}^{(1,1)}=0, \\
\medskip
\disp 
R_{m-1,n}^{(1,0)}R_{m,n-1}^{(0,0)}
-(x+1)R_{m-1,n-1}^{(1,1)}R_{m,n}^{(0,-1)}
+xR_{m,n-1}^{(1,0)}R_{m-1,n}^{(0,0)}=0, \\
\medskip
\disp 
(x+1)R_{m,n}^{(1,1)}R_{m-1,n-1}^{(0,-1)}
-xR_{m-1,n}^{(1,0)}R_{m,n-1}^{(0,0)}
-R_{m,n-1}^{(1,0)}R_{m-1,n}^{(0,0)}=0, 
\end{array}
\end{equation}
\begin{equation}
\begin{array}{l}
\medskip
\disp 
(2c-d-m-n)R_{m-1,n}^{(0,0)}R_{m,n}^{(0,-1)}
+c(x-1)R_{m-1,n}^{(1,0)}R_{m,n}^{(-1,-1)}
+(2n+1)R_{m,n+1}^{(0,0)}R_{m-1,n-1}^{(0,-1)}=0, \\
\medskip
\disp 
(2c-d-m-n)R_{m-1,n}^{(0,0)}R_{m,n-1}^{(0,0)}
-cR_{m-1,n-1}^{(1,1)}R_{m,n}^{(-1,-1)}
-(c-d)R_{m,n}^{(0,1)}R_{m-1,n-1}^{(0,-1)}=0, \\
\medskip
\disp 
(2c-d-m-n)xR_{m,n}^{(0,-1)}R_{m,n-1}^{(0,0)}
-c(x-1)R_{m,n-1}^{(1,0)}R_{m,n}^{(-1,-1)}
+(2m+1)xR_{m+1,n}^{(0,0)}R_{m-1,n-1}^{(0,-1)}=0, 
\end{array}
\end{equation}
\begin{equation}
\begin{array}{l}
\medskip
\disp 
(d+m-n)R_{m-1,n}^{(0,1)}R_{m,n}^{(0,0)}
-c(x+1)R_{m,n}^{(1,2)}R_{m-1,n}^{(-1,-1)}
-(2n+1)R_{m-1,n+1}^{(0,0)}R_{m,n-1}^{(0,1)}=0, \\
\medskip
\disp 
(d+m-n)xR_{m,n}^{(0,0)}R_{m,n-1}^{(0,-1)}
-c(x+1)R_{m,n-1}^{(1,1)}R_{m,n}^{(-1,-2)}
-(2m+1)xR_{m-1,n}^{(0,-1)}R_{m+1,n-1}^{(0,0)}=0, \\
\medskip
\disp 
(d+m-n)R_{m,n}^{(0,0)}R_{m-1,n-1}^{(0,0)}
-cR_{m,n-1}^{(1,1)}R_{m-1,n}^{(-1,-1)}
+(c-d) R_{m-1,n}^{(0,-1)}R_{m,n-1}^{(0,1)}=0, 
\end{array}
\end{equation}
\begin{equation}
\begin{array}{l}
\medskip
\disp 
(2c-d+m+n)xR_{m,n}^{(0,1)}R_{m,n-1}^{(0,0)}
-c(x-1)R_{m,n}^{(1,1)}R_{m,n-1}^{(-1,0)}
+(2m+1)xR_{m-1,n-1}^{(0,1)}R_{m+1,n}^{(0,0)}=0, \\
\medskip
\disp 
(2c-d+m+n)R_{m-1,n}^{(0,0)}R_{m,n-1}^{(0,0)}
-cR_{m,n}^{(1,1)}R_{m-1,n-1}^{(-1,-1)}
-(c-d)R_{m-1,n-1}^{(0,1)}R_{m,n}^{(0,-1)}=0, \\
\medskip
\disp 
(2c-d+m+n)R_{m,n}^{(0,1)}R_{m-1,n}^{(0,0)}
+c(x-1)R_{m,n}^{(1,1)}R_{m-1,n}^{(-1,0)}
+(2n+1)R_{m-1,n-1}^{(0,1)}R_{m,n+1}^{(0,0)}=0, 
\end{array}
\end{equation}
\begin{equation}
\begin{array}{l}
\medskip
\disp 
(d-m+n)R_{m,n}^{(0,0)}R_{m-1,n-1}^{(0,0)}
-cR_{m-1,n}^{(1,1)}R_{m,n-1}^{(-1,-1)}
+(c-d)R_{m,n-1}^{(0,-1)}R_{m-1,n}^{(0,1)}=0, \\
\medskip
\disp 
(d-m+n)xR_{m,n}^{(0,0)}R_{m,n-1}^{(0,1)}
-c(x+1)R_{m,n}^{(1,2)}R_{m,n-1}^{(-1,-1)}
-(2m+1)xR_{m+1,n-1}^{(0,0)}R_{m-1,n}^{(0,1)}=0, \\
\medskip
\disp 
(d-m+n)R_{m,n}^{(0,0)}R_{m-1,n}^{(0,-1)}
-c(x+1)R_{m-1,n}^{(1,1)}R_{m,n}^{(-1,-2)}
-(2n+1)R_{m,n-1}^{(0,-1)}R_{m-1,n+1}^{(0,0)}=0, 
\end{array}
\end{equation}
\begin{equation}
\begin{array}{l}
\medskip
\disp 
2R_{m-1,n}^{(0,0)}R_{m,n}^{(0,-1)}
-(x+1)R_{m,n}^{(1,1)}R_{m-1,n}^{(-1,-2)}
+(x-1)R_{m-1,n}^{(1,0)}R_{m,n}^{(-1,-1)}=0, \\
\medskip
\disp 
2xR_{m,n}^{(0,0)}R_{m,n-1}^{(0,-1)}
-(x+1)R_{m,n-1}^{(1,1)}R_{m,n}^{(-1,-2)}
-(x-1)R_{m,n}^{(1,0)}R_{m,n-1}^{(-1,-1)}=0, \\
\medskip
\disp 
2R_{m,n}^{(0,0)}R_{m-1,n-1}^{(0,0)}
-R_{m,n-1}^{(1,1)}R_{m-1,n}^{(-1,-1)}
-R_{m-1,n}^{(1,1)}R_{m,n-1}^{(-1,-1)}=0, \\
\medskip
\disp 
2R_{m,n}^{(0,0)}R_{m-1,n}^{(0,-1)}
-(x+1)R_{m-1,n}^{(1,1)}R_{m,n}^{(-1,-2)}
+(x-1)R_{m,n}^{(1,0)}R_{m-1,n}^{(-1,-1)}=0, \\
\medskip
\disp 
2xR_{m,n}^{(0,-1)}R_{m,n-1}^{(0,0)}
-(x+1)R_{m,n}^{(1,1)}R_{m,n-1}^{(-1,-2)}
-(x-1)R_{m,n-1}^{(1,0)}R_{m,n}^{(-1,-1)}=0, \\
\disp 
2R_{m-1,n}^{(0,0)}R_{m,n-1}^{(0,0)}
-R_{m,n}^{(1,1)}R_{m-1,n-1}^{(-1,-1)}
-R_{m-1,n-1}^{(1,1)}R_{m,n}^{(-1,-1)}=0, 
\end{array}
\end{equation}
where we denote as 
\begin{equation}
R_{m,n}^{(i,j)}=R_{m,n}(c+i,d+j). 
\end{equation}
\end{prop}
From the above discussion, 
now the proof of Theorem \ref{detV} is reduced to that of Proposition
\ref{bi:rel:V}. 

It is possible to reduce the number of bilinear relations to be proved
by the following symmetries of $R_{m,n}$. 

\begin{lem}\label{simple:V}
We have the relations for $m,n \in \BZ_{\ge 0}$ 
\begin{equation}
R_{n,m}(x^{-1})=R_{m,n}(x), \label{x->1/x:V}
\end{equation}
\begin{equation}
R_{m,n}(-c,-d)=(-1)^{m(m+1)/2+n(n+1)/2}R_{m,n}(c,d), \label{cd:V}
\end{equation}
\begin{equation}
R_{m,n}(-x;c,2c-d)=(-1)^{m(m+1)/2+n(n+1)/2}R_{m,n}(x;c,d). 
\label{d:V}
\end{equation}
\end{lem}

\noindent
{\it Proof.}\quad 
The first relation (\ref{x->1/x:V}) is easily obtained from the
definition (\ref{def:pq:V}) and (\ref{P6:alg:tau}). 
To verify the second relation (\ref{cd:V}), we introduce two sets of
polynomials $\bar{p}_k=\bar{p}_k^{(c,d)}(x)$ and 
$\bar{q}_k=\bar{q}_k^{(c,d)}(x),~k \in \BZ$, by 
\begin{equation}
\begin{array}{c}
\disp 
\sum_{k=0}^{\infty}\bar{p}_k^{(c,d)}(x)\lambda^k=G(x;-c,-d;-\lambda), \quad 
\bar{p}_k^{(c,d)}(x)=0\ \mbox{for}\ k<0, \\
\disp 
\bar{q}_k^{(c,d)}(x)=\bar{p}_k^{(c,d)}(x^{-1}), 
\end{array}
\end{equation}
where $G$ is the generating function (\ref{GF:V}). 
Since we have 
\begin{equation}
\frac{G(x;-c,-d;-\lambda)}{G(x;c,d;\lambda)}=
(1-\lambda^2)^{-c+d}\left(1-x^2\lambda^2\right)^c, 
\end{equation}
we see that 
\begin{equation}
\bar{p}_k(x)=p_k(x)+\sum_{j=1}^{\infty}\rho_j(x)p_{k-2j}(x), \quad 
\bar{q}_k(x)=q_k(x)+\sum_{j=1}^{\infty}\rho_j(x^{-1})q_{k-2j}(x),
\end{equation}
where $\rho_j(x)=\rho_j(x;c,d)$ are some functions. 
Therefore, $R_{m,n}$ for $m,n \in \BZ_{\ge 0}$ can be expressed in terms
of the same determinant as (\ref{P6:alg:tau}) with the entries $p_k$ and
$q_k$ replaced by $\bar{p}_k$ and $\bar{q}_k$, respectively. 
Noticing that 
\begin{equation}
\bar{p}_k^{(c,d)}(x)=(-1)^k p_k^{(-c,-d)}(x), \quad 
\bar{q}_k^{(c,d)}(x)=(-1)^k q_k^{(-c,-d)}(x), 
\end{equation}
we obtain the relation (\ref{cd:V}). 
The third relation (\ref{d:V}) is verified similarly. 
\hfill\qed

\medskip

By the symmetries of $R_{m,n}$ described by (\ref{neg:R}) and Lemma
\ref{simple:V}, it is sufficient to prove the following bilinear
relations for $m,n \in \BZ_{\ge 0}$, 
\begin{equation}
(x+1)R_{m,n}^{(1,1)}R_{m-1,n-1}^{(0,-1)}
-xR_{m-1,n}^{(1,0)}R_{m,n-1}^{(0,0)}
-R_{m,n-1}^{(1,0)}R_{m-1,n}^{(0,0)}=0, \label{bi:norm:V}
\end{equation}
\begin{equation}
(d+m-n)R_{m-1,n}^{(0,1)}R_{m,n}^{(0,0)}
-c(x+1)R_{m,n}^{(1,2)}R_{m-1,n}^{(-1,-1)}
-(2n+1)R_{m-1,n+1}^{(0,0)}R_{m,n-1}^{(0,1)}=0, \label{bi:BT:e1:V}
\end{equation}
\begin{equation}
(d+m-n)R_{m,n}^{(0,0)}R_{m-1,n-1}^{(0,0)}
-cR_{m,n-1}^{(1,1)}R_{m-1,n}^{(-1,-1)}
+(c-d) R_{m-1,n}^{(0,-1)}R_{m,n-1}^{(0,1)}=0, \label{bi:BT:e2:V}
\end{equation}
\begin{equation}
2R_{m,n}^{(0,0)}R_{m-1,n-1}^{(0,0)}
-R_{m,n-1}^{(1,1)}R_{m-1,n}^{(-1,-1)}
-R_{m-1,n}^{(1,1)}R_{m,n-1}^{(-1,-1)}=0, \label{bi:BT:c1:V}
\end{equation}
\begin{equation}
2R_{m-1,n}^{(0,0)}R_{m,n-1}^{(0,0)}
-R_{m,n}^{(1,1)}R_{m-1,n-1}^{(-1,-1)}
-R_{m-1,n-1}^{(1,1)}R_{m,n}^{(-1,-1)}=0, \label{bi:BT:c2:V}
\end{equation}
\begin{equation}
2R_{m-1,n}^{(0,0)}R_{m,n}^{(0,-1)}
-(x+1)R_{m,n}^{(1,1)}R_{m-1,n}^{(-1,-2)}
+(x-1)R_{m-1,n}^{(1,0)}R_{m,n}^{(-1,-1)}=0. \label{bi:BT:c0:V}
\end{equation}

From the symmetry (\ref{d:V}) and the bilinear relation (\ref{bi:norm:V}), we have 
\begin{equation}
(x-1)R_{m,n}^{(0,0)}R_{m-1,n-1}^{(1,0)}
-xR_{m-1,n}^{(0,-1)}R_{m,n-1}^{(1,1)}
+R_{m,n-1}^{(0,-1)}R_{m-1,n}^{(1,1)}=0. \label{bi:norm':V}
\end{equation}
Then, it is possible to derive the bilinear relation (\ref{bi:BT:c0:V}) as follows, 
\begin{equation}
R_{m-1,n}^{(0,0)}\times (\ref{bi:BT:c1:V})|_{d \to d-1}
-R_{m-1,n}^{(-1,-2)}\times (\ref{bi:norm:V})
+R_{m-1,n}^{(1,0)}\times (\ref{bi:norm':V})|_{c \to c-1,d \to d-1}
=R_{m-1,n-1}^{(0,-1)}\times (\ref{bi:BT:c0:V}). 
\end{equation}
Therefore, the bilinear relations to be proved are
(\ref{bi:norm:V})-(\ref{bi:BT:c2:V}). 

In the following, we show that these bilinear relations
(\ref{bi:norm:V})-(\ref{bi:BT:c2:V}) are reduced to Jacobi's identity of
determinants. 
Let $D$ be an $(m+n+1) \times (m+n+1)$ determinant and 
$\disp 
D\left[\begin{array}{cccc}
        i_1 & i_2 & \cdots & i_k \\ 
        j_1 & j_2 & \cdots & j_k
       \end{array}
 \right]$ 
the minor that are obtained by deleting the rows with indices 
$i_1,\ldots,i_k$ and the columns with indices $j_1,\ldots,j_k$. 
Then, we have Jacobi's identities 
\begin{equation}
D \cdot D\left[\begin{array}{cc}
                m &  m+1  \\
                1 & m+n+1
              \end{array}
         \right]=
D\left[\begin{array}{c}
        m \\
        1
       \end{array}
 \right]
D\left[\begin{array}{c}
         m+1  \\
        m+n+1
       \end{array}
 \right]-
D\left[\begin{array}{c}
          m \\
        m+n+1
       \end{array}
 \right]
D\left[\begin{array}{c}
        m+1   \\
         1
       \end{array}
 \right],     \label{Jacobi:1}
\end{equation}
\begin{equation}
D \cdot D\left[\begin{array}{cc}
                m &  m+1  \\
                1  &  2
              \end{array}
         \right]=
D\left[\begin{array}{c}
        m \\
        1
       \end{array}
 \right]
D\left[\begin{array}{c}
        m+1  \\
         2
       \end{array}
 \right]-
D\left[\begin{array}{c}
        m   \\
        2
       \end{array}
 \right]
D\left[\begin{array}{c}
        m+1 \\
         1
       \end{array}
 \right],     \label{Jacobi:2}
\end{equation}
\begin{equation}
D \cdot D\left[\begin{array}{cc}
                1 &  m+1  \\
                2 & m+n+1
              \end{array}
         \right]=
D\left[\begin{array}{c}
        1 \\
        2
       \end{array}
 \right]
D\left[\begin{array}{c}
         m+1  \\
        m+n+1
       \end{array}
 \right]-
D\left[\begin{array}{c}
          1 \\
        m+n+1
       \end{array}
 \right]
D\left[\begin{array}{c}
        m+1   \\
         2
       \end{array}
 \right].     \label{Jacobi:3}
\end{equation}

First, we give the proof of the bilinear relations
(\ref{bi:norm:V})-(\ref{bi:BT:e2:V}). 
We have the following lemmas. 
\begin{lem}\label{V1}
Put 
\begin{equation}
D = 
 \left|
  \begin{array}{ccccc}
   -q_1^{(c-m-n,d-m-n)}      & q_1^{(c-m-n+1,d-m-n)}      & \cdots 
                             & q_{-m-n+3}^{(c-1,d-2)}     & q_{-m-n+2}^{(c,d-1)}\\
   -q_3^{(c-m-n,d-m-n)}      & q_3^{(c-m-n+1,d-m-n)}      & \cdots 
                             & q_{-m-n+5}^{(c-1,d-2)}     & q_{-m-n+4}^{(c,d-1)}\\
   \vdots & \vdots & \ddots & \vdots & \vdots \\
   -q_{2m-1}^{(c-m-n,d-m-n)} & q_{2m-1}^{(c-m-n+1,d-m-n)} & \cdots 
                             & q_{m-n+1}^{(c-1,d-2)}      & q_{m-n}^{(c,d-1)}   \\
   x^{-m-n}p_{2n}^{(c-m-n,d-m-n)} & x^{-m-n+1}p_{2n}^{(c-m-n+1,d-m-n)} & \cdots 
   & x^{-1}p_{2n}^{(c-1,d-2)}     & p_{2n}^{(c,d-1)} \\
   \vdots & \vdots & \ddots & \vdots & \vdots     \\
   x^{-m-n}p_2^{(c-m-n,d-m-n)}    & x^{-m-n+1}p_2^{(c-m-n+1,d-m-n)}    & \cdots 
   & x^{-1}p_2^{(c-1,d-2)}        & p_2^{(c,d-1)}  \\
   x^{-m-n}p_0^{(c-m-n,d-m-n)}    & x^{-m-n+1}p_0^{(c-m-n+1,d-m-n)}    & \cdots 
   & x^{-1}p_0^{(c-1,d-2)}        & p_0^{(c,d-1)}
  \end{array}
 \right|. 
\end{equation}
Then, we have 
\begin{equation}
\begin{array}{c}
\medskip
\disp
 D=(-1)^m(1+x^{-1})^{m+n}R_{m,n}^{(0,0)}, \quad 
 D\left[\begin{array}{cc}
         m &  m+1  \\
         1 & m+n+1
        \end{array}
  \right]
 =x^{-n}R_{m-1,n-1}^{(-1,-2)}, \\
\medskip
\disp
 D\left[\begin{array}{c}
          m \\
          1
        \end{array}
  \right]
 =R_{m-1,n}^{(0,-1)}, \quad 
 D\left[\begin{array}{c}
          m+1 \\
         m+n+1
        \end{array}
  \right]
 =(-1)^m x^{-n}(1+x^{-1})^{m+n-1}R_{m,n-1}^{(-1,-1)}, \\
\disp
 D\left[\begin{array}{c}
           m \\
         m+n+1
        \end{array}
  \right]
 =(-1)^{m-1}x^{-n-1}(1+x^{-1})^{m+n-1}R_{m-1,n}^{(-1,-1)}, \quad 
 D\left[\begin{array}{c}
          m+1  \\
           1
        \end{array}
  \right]
 =R_{m,n-1}^{(0,-1)}. 
\end{array}
\end{equation}
\end{lem}
\begin{lem}\label{V2}
Put 
\begin{equation}
D = 
 \left|
  \begin{array}{ccccc}
   \widetilde{q}_1^{(c-m-n,d-m-n)}      & q_1^{(c-m-n,d-m-n-1)}      & \cdots 
            & q_{-m-n+3}^{(c-2,d-3)}     & q_{-m-n+2}^{(c-1,d-2)}\\
   \widetilde{q}_3^{(c-m-n,d-m-n)}      & q_3^{(c-m-n,d-m-n-1)}      & \cdots 
            & q_{-m-n+5}^{(c-2,d-3)}     & q_{-m-n+4}^{(c-1,d-2)}\\
   \vdots & \vdots & \ddots & \vdots & \vdots \\
   \widetilde{q}_{2m-1}^{(c-m-n,d-m-n)} & q_{2m-1}^{(c-m-n,d-m-n-1)} & \cdots 
            & q_{m-n+1}^{(c-2,d-3)}      & q_{m-n}^{(c-1,d-2)}   \\
   \widehat{p}_{2n}^{(c-m-n,d-m-n)} & x^{-m-n+1}p_{2n+1}^{(c-m-n,d-m-n-1)} 
    & \cdots                        & x^{-1}p_{2n+1}^{(c-2,d-3)} & p_{2n+1}^{(c-1,d-2)} \\
   \vdots & \vdots & \ddots & \vdots & \vdots     \\
   \widehat{p}_2^{(c-m-n,d-m-n)}    & x^{-m-n+1}p_3^{(c-m-n,d-m-n-1)}
    & \cdots                        & x^{-1}p_3^{(c-2,d-3)}      & p_3^{(c-1,d-2)}      \\
   \widehat{p}_0^{(c-m-n,d-m-n)}    & x^{-m-n+1}p_1^{(c-m-n,d-m-n-1)}
   & \cdots                         & x^{-1}p_1^{(c-2,d-3)}      & p_1^{(c-1,d-2)}
  \end{array}
 \right|, 
\end{equation}
with 
\begin{equation}
\widehat{p}_{2k}^{(c-m-n,d-m-n)}=\frac{x^{-m-n}}{2k+1}p_{2k}^{(c-m-n,d-m-n)}, \quad 
\widetilde{q}_{2k-1}^{(c-m-n,d-m-n)}=\frac{q_{2k-1}^{(c-m-n,d-m-n)}}{(d-m-n+2k-2)x}. 
\end{equation}
Then, we have 
\begin{equation}
\begin{array}{c}
\medskip
\disp
 D=(-1)^{m+n}(1+x)^{m+n}x^{-m}
   \frac{\disp \prod_{j=1}^{m+n}(c-m-n+j-1)}
        {\disp \prod_{i=1}^m(d-m-n+2i-2)\prod_{k=0}^n(2k+1)}
   R_{m,n}^{(0,0)}, \\ 
\medskip
\disp
 D\left[\begin{array}{c}
           m \\
         m+n+1
        \end{array}
  \right]
 =(-1)^{m+n-1}(1+x)^{m+n-1}x^{-m-n}
   \frac{\disp \prod_{j=1}^{m+n-1}(c-m-n+j-1)}
        {\disp \prod_{i=1}^{m-1}(d-m-n+2i-2)\prod_{k=0}^n(2k+1)}
   R_{m-1,n}^{(-1,-1)}, \\
\medskip
\disp
 D\left[\begin{array}{c}
          m+1 \\
         m+n+1
        \end{array}
  \right]
 =(-1)^{m+n-1}(1+x)^{m+n-1}x^{-m-n}
   \frac{\disp \prod_{j=1}^{m+n-1}(c-m-n+j-1)}
        {\disp \prod_{i=1}^m(d-m-n+2i-2)\prod_{k=0}^{n-1}(2k+1)}
   R_{m,n-1}^{(-1,-1)}, \\
\disp
 D\left[\begin{array}{c}
          m \\
          1
        \end{array}
  \right]
 =R_{m-1,n+1}^{(-1,-2)}, \quad 
 D\left[\begin{array}{c}
          m+1  \\
           1
        \end{array}
  \right]
 =R_{m,n}^{(-1,-2)}, \quad 
 D\left[\begin{array}{cc}
         m &  m+1  \\
         1 & m+n+1
        \end{array}
  \right]
 =x^{-n}R_{m-1,n}^{(-2,-3)}. 
\end{array}
\end{equation}
\end{lem}
\begin{lem}\label{V3}
Put 
\begin{equation}
D= 
 \left|
  \begin{array}{ccccc}
   \widetilde{q}_1^{(c,d-m-n)}      & q_1^{(c-1,d-m-n-1)}      & \cdots 
            & q_{-m-n+3}^{(c-1,d-3)}     & q_{-m-n+2}^{(c-1,d-2)}\\
   \widetilde{q}_3^{(c,d-m-n)}      & q_3^{(c-1,d-m-n-1)}      & \cdots 
            & q_{-m-n+5}^{(c-1,d-3)}     & q_{-m-n+4}^{(c-1,d-2)}\\
   \vdots & \vdots & \ddots & \vdots & \vdots \\
   \widetilde{q}_{2m-1}^{(c,d-m-n)} & q_{2m-1}^{(c-1,d-m-n-1)} & \cdots 
            & q_{m-n+1}^{(c-1,d-3)}      & q_{m-n}^{(c-1,d-2)}   \\
   \widehat{p}_{2n}^{(c,d-m-n)} & (-1)^{m+n-1}p_{2n+1}^{(c-1,d-m-n-1)} 
    & \cdots                        & (-1)^1 p_{2n+1}^{(c-1,d-3)} & p_{2n+1}^{(c-1,d-2)} \\
   \vdots & \vdots & \ddots & \vdots & \vdots     \\
   \widehat{p}_2^{(c,d-m-n)}    & (-1)^{m+n-1}p_3^{(c-1,d-m-n-1)}
    & \cdots                        & (-1)^1 p_3^{(c-1,d-3)}      & p_3^{(c-1,d-2)}      \\
   \widehat{p}_0^{(c,d-m-n)}    & (-1)^{m+n-1}p_1^{(c-1,d-m-n-1)}
   & \cdots                         & (-1)^1 p_1^{(c-1,d-3)}      & p_1^{(c-1,d-2)}
  \end{array}
 \right|, 
\end{equation}
with 
\begin{equation}
\widehat{p}_{2k}^{(c,d-m-n)}=(-1)^{m+n}\frac{p_{2k}^{(c,d-m-n)}}{2k+1}, \quad 
\widetilde{q}_{2k-1}^{(c,d-m-n)}=\frac{q_{2k-1}^{(c,d-m-n)}}{(d-m-n+2k-2)x}. 
\end{equation}
Then, we have 
\begin{eqnarray}
&& \hskip-20pt 
D=(-1)^{m+n}(1+x)^{m+n}x^{-m}
   \frac{\disp \prod_{j=1}^{m+n}(c-d+m+n-j+1)}
        {\disp \prod_{i=1}^m(d-m-n+2i-2)\prod_{k=0}^n(2k+1)}
   R_{m,n}^{(0,0)}, \cr 
&& \hskip-20pt 
 D\left[\begin{array}{c}
           m \\
         m+n+1
        \end{array}
  \right]
 =(-1)^m(1+x)^{m+n-1}x^{-m+1}
   \frac{\disp \prod_{j=1}^{m+n-1}(c-d+m+n-j+1)}
        {\disp \prod_{i=1}^{m-1}(d-m-n+2i-2)\prod_{k=0}^n(2k+1)}
   R_{m-1,n}^{(0,-1)}, \cr
&& \hskip-20pt 
 D\left[\begin{array}{c}
          m+1 \\
         m+n+1
        \end{array}
  \right]
 =(-1)^{m-1}(1+x)^{m+n-1}x^{-m}
   \frac{\disp \prod_{j=1}^{m+n-1}(c-d+m+n-j+1)}
        {\disp \prod_{i=1}^m(d-m-n+2i-2)\prod_{k=0}^{n-1}(2k+1)}
   R_{m,n-1}^{(0,-1)}, \cr 
&& \hskip-20pt 
 D\left[\begin{array}{c}
          m \\
          1
        \end{array}
  \right]
 =R_{m-1,n+1}^{(-1,-2)}, \quad 
 D\left[\begin{array}{c}
          m+1  \\
           1
        \end{array}
  \right]
 =R_{m,n}^{(-1,-2)}, \quad 
 D\left[\begin{array}{cc}
         m &  m+1  \\
         1 & m+n+1
        \end{array}
  \right]
 =(-1)^n R_{m-1,n}^{(-1,-3)}. 
\end{eqnarray}
\end{lem}
It is easy to see that the bilinear relations (\ref{bi:norm:V}) and
(\ref{bi:BT:e1:V}) follow immediately from Jacobi's identity
(\ref{Jacobi:1}) by using Lemma \ref{V1} and \ref{V2}, respectively. 
By the Lemma \ref{V3}, Jacobi's identity (\ref{Jacobi:1}) is reduced to 
\begin{equation}
(d+m-n)xR_{m-1,n}^{(1,1)}R_{m,n}^{(0,0)}
+(c-d)(x+1)R_{m,n}^{(1,2)}R_{m-1,n}^{(0,-1)}
+(2n+1)R_{m-1,n+1}^{(0,0)}R_{m,n-1}^{(1,1)}=0. \label{bi:V}
\end{equation}
Then, the bilinear relation (\ref{bi:BT:e2:V}) is derived as follows, 
\begin{equation}
R_{m,n-1}^{(0,1)}\times (\ref{bi:V})
+R_{m,n-1}^{(1,1)}\times (\ref{bi:BT:e1:V})
+(d+m-n)R_{m,n}^{(0,0)}\times (\ref{bi:norm:V})|_{d \to d+1}
=(x+1)R_{m,n}^{(1,2)}\times (\ref{bi:BT:e2:V}). 
\end{equation}
The proof of Lemmas \ref{V1}$\sim$\ref{V3} is given in Appendix \ref{PoL}. 

Next, we prove the bilinear relations (\ref{bi:BT:c1:V}) and
(\ref{bi:BT:c2:V}). 
We have the following lemmas. 
\begin{lem}\label{V4}
Put 
\begin{equation}
D = 
 \left|
  \begin{array}{ccccc}
  -x^{-1}q_1^-                              & x^{-1}q_1^+      & q_1     & \cdots & q_{-m-n+4}\\
  -x^{-1}(q_3^-+x^{-2}q_1^-)                & x^{-1}q_3^+      & q_3     & \cdots & q_{-m-n+6}\\
  \vdots                                    & \vdots           & \vdots  & \ddots & \vdots    \\
  -x^{-1}(q_{2m-1}^-+\cdots +x^{-2m+2}q_1^-)& x^{-1}q_{2m-1}^+ & q_{2m-1}& \cdots & q_{m-n+2} \\
  p_{n-m+1}^-+\cdots +x^{2n-2}p_{-n-m+3}^- & p_{n-m+1}^+  & p_{n-m+2} & \cdots & p_{2n-1}\\
  \vdots                                   & \vdots       & \vdots    & \ddots & \vdots  \\
  p_{-n-m+5}^-+x^2p_{-n-m+3}^-           & p_{-n-m+5}^+ & p_{-n-m+6}& \cdots & p_3     \\
  p_{-n-m+3}^-                           & p_{-n-m+3}^+ & p_{-n-m+4}& \cdots & p_1
  \end{array}
 \right|, 
\end{equation}
with $p_k^{\pm}=p_k^{(c\pm 1,d\pm 1)}$. 
Then, we have 
\begin{equation}
\begin{array}{c}
\medskip
\disp
 D\left[\begin{array}{c}
          m \\
          1
        \end{array}
  \right]
 =x^{-m+1}R_{m-1,n}^{(1,1)}, \quad 
 D\left[\begin{array}{c}
          m+1 \\
           1
        \end{array}
  \right]
 =x^{-m}R_{m,n-1}^{(1,1)}, \\
\medskip
\disp
 D\left[\begin{array}{c}
          m \\
          2
        \end{array}
  \right]
 =(-x)^{-m+1}R_{m-1,n}^{(-1,-1)}, \quad 
 D\left[\begin{array}{c}
          m+1  \\
           2
        \end{array}
  \right]
 =(-x)^{-m}R_{m,n-1}^{(-1,-1)}, \\
\disp
 D\left[\begin{array}{cc}
           m &  m+1  \\
           1  &  2
        \end{array}
  \right]
 =R_{m-1,n-1}^{(0,0)}, 
\end{array}
\end{equation}
and 
\begin{equation}
D=2(-1)^{-m}x^{-2m+1}R_{m,n}^{(0,0)}. 
\end{equation}
\end{lem}
\begin{lem}\label{V5}
Put 
\begin{equation}
D = 
 \left|
  \begin{array}{ccccc}
    1 & 0 & 0 & \cdots & 0 \\
   -x^{-1}q_1^-                            & x^{-1}q_1^+      & q_1     & \cdots & q_{-m-n+3}\\
   -x^{-1}(q_3^-+x^{-2}q_1^-)              & x^{-1}q_3^+      & q_3     & \cdots & q_{-m-n+5}\\
   \vdots                                  & \vdots           & \vdots  & \ddots & \vdots    \\
   -x^{-1}(q_{2m-1}^-+\cdots +x^{-2m+2}q_1^-)& x^{-1}q_{2m-1}^+ & q_{2m-1}& \cdots & q_{m-n+1} \\
   p_{n-m}^-+\cdots +x^{2n-2}p_{-n-m+2}^- & p_{n-m}^+    & p_{n-m+1} & \cdots & p_{2n-1}\\
   \vdots                             & \vdots       & \vdots    & \ddots & \vdots  \\
   p_{-n-m+4}^-+x^2p_{-n-m+2}^-       & p_{-n-m+4}^+ & p_{-n-m+5}& \cdots & p_3     \\
   p_{-n-m+2}^-                       & p_{-n-m+2}^+ & p_{-n-m+3}& \cdots & p_1
  \end{array}
 \right|. 
\end{equation}
Then, we have 
\begin{equation}
\begin{array}{c}
\medskip
\disp
 D=x^{-m}R_{m,n}^{(1,1)}, \quad 
 D\left[\begin{array}{cc}
         1 &  m+1  \\
         2 & m+n+1
        \end{array}
  \right]
 =(-x)^{-m+1}R_{m-1,n-1}^{(-1,-1)},  \\
\medskip
\disp
 D\left[\begin{array}{c}
          1 \\
          2
        \end{array}
  \right]
 =(-x)^{-m}R_{m,n}^{(-1,-1)}, \quad 
 D\left[\begin{array}{c}
          m+1 \\
         m+n+1
        \end{array}
  \right]
 =x^{-m+1}R_{m-1,n-1}^{(1,1)}, \\
\disp
 D\left[\begin{array}{c}
          m+1 \\
           2
        \end{array}
  \right]
 =R_{m-1,n}^{(0,0)}, \quad 
 D\left[\begin{array}{c}
           1  \\
         m+n+1
        \end{array}
  \right]
 =2(-1)^{-m}x^{-2m+1}R_{m,n-1}^{(0,0)}. 
\end{array}
\end{equation}
\end{lem}
From Lemma \ref{V4}, Jacobi's identity (\ref{Jacobi:2}) is lead to the
bilinear relation (\ref{bi:BT:c1:V}). 
Lemma \ref{V5} and Jacobi's identity (\ref{Jacobi:3}) give the bilinear
relation (\ref{bi:BT:c2:V}). 
We also give the proof of Lemma \ref{V4} and \ref{V5} in Appendix
\ref{PoL}.

\section{Degeneration of algebraic solutions \label{degen}}
It is well known that, starting from P$_{\rm VI}$, one can obtain
P$_{\rm V}, \ldots$, P$_{\rm I}$ by successive limiting procedures in
the following diagram~\cite{Pa,GtoP}, 
\begin{equation}
\begin{array}{c}
\mbox{P}_{\rm VI} \lora \mbox{P}_{\rm V}  \lora \mbox{P}_{\rm III} \\
\hskip30pt                \da        \hskip30pt   \da              \\
\hskip71pt              \mbox{P}_{\rm IV} \lora \mbox{P}_{\rm II}  \lora \mbox{P}_{\rm I}. 
\end{array}
\end{equation}
It is also known that each Painlev\'e equation, except for P$_{\rm I}$,
admits particular solutions expressed by special functions, 
and that the coalescence diagram of these special functions is given as 
\begin{equation}
\begin{array}{c}
\mbox{hypergeometric} \lora \mbox{confluent hypergeometric} \lora \mbox{Bessel} \\
\hskip120pt \da \hskip100pt \da \\
\hskip100pt \mbox{Hermite-Weber} \hskip20pt \lora \hskip20pt \mbox{Airy}. 
\end{array}
\end{equation}

How is the degeneration diagram of algebraic (or rational) solutions
that originate from the fixed points of Dynkin automorphisms?
In this section, we show that, starting from the family of algebraic
solutions to P$_{\rm VI}$ given in Section \ref{main}, 
we can obtain that of the rational solutions to P$_{\rm V}$,~P$_{\rm
III}$ and P$_{\rm II}$ by degeneration in the following diagram, 
\begin{equation}
\begin{array}{c}
          \mbox{P}_{\rm VI} \lora \mbox{P}_{\rm V}  \\
\da \hskip28pt \da \\
\hskip3pt \mbox{P}_{\rm III}\lora \mbox{P}_{\rm II}. 
\end{array}   \label{coal:alg}
\end{equation}
The horizontal arrows in (\ref{coal:alg}) are naturally obtained as the
coalescence of Painlev\'e equations (or corresponding Hamilton
systems). 
We need to remark on the vertical arrows. 
The degeneration procedures themselves are obtained as the combination
of those for equations or Hamilton systems. 
Nevertheless, the degeneration of the algebraic (or rational) solutions 
seems to be direct one. 
Namely, we believe that the rational solutions of P$_{\rm V}$ cannot
degenerate to those of P$_{\rm III}$ and that the rational solutions of
P$_{\rm IV}$ expressed in terms of Okamoto polynomials cannot be linked
with the above diagram.

\subsection{Degeneration from P$_{\rm VI}$ to P$_{\rm V}$}
As is known~\cite{O2}, P$_{\rm V}$ 
\begin{equation}
\frac{d^2y}{dt^2}=
\left(\frac{1}{2y}+\frac{1}{y-1}\right)\left(\frac{dy}{dt}\right)^2-\frac{1}{t}\frac{dy}{dt}
+\frac{(y-1)^2}{2t^2}\left(\kappa_{\infty}^2 y-\frac{\kappa_0^2}{y}\right)
-(\theta+1)\frac{y}{t}-\frac{y(y+1)}{2(y-1)},  \label{P5}
\end{equation}
is equivalent to the Hamilton system 
\begin{equation}
\hskip-40pt
\mbox{S$_{\rm V}$~:} \hskip30pt
q'=\frac{\partial H}{\partial p}, \quad p'=-\frac{\partial H}{\partial q}, \quad
'=t\frac{d}{dt},  \label{cano5}
\end{equation}
with the Hamiltonian 
\begin{equation}
\begin{array}{c}
\smallskip
\disp 
H=q(q-1)^2p^2-\left[\kappa_0(q-1)^2+\theta q(q-1)+tq\right]p+\kappa(q-1), \\
\disp 
\kappa=\frac{1}{4}(\kappa_0+\theta)^2-\frac{1}{4}\kappa_{\infty}^2. 
\end{array}   \label{H5}
\end{equation}
This system can be derived from S$_{\rm VI}$ by degeneration~\cite{GtoP}. 
The Hamilton equation (\ref{cano6}) with the Hamiltonian (\ref{H6}) is
reduced to (\ref{cano5}) with (\ref{H5}) by putting 
\begin{equation}
t \to 1-\ep t, \quad \kappa_1 \to \ep^{-1}+\theta+1, \quad \theta \to -\ep^{-1}, 
\end{equation}
and taking the limit of $\ep \to 0$. 

The rational solutions of S$_{\rm V}$ are expressed as
follows~\cite{P5:rat}. 

\begin{prop}\label{S5:rat}
Let $p_k=p_k^{(r)}(z)$ and $q_k=q_k^{(r)}(z),~k \in \BZ$, be two sets of
 polynomials defined by 
\begin{equation}
\begin{array}{c}
\disp 
 \sum_{k=0}^{\infty}p_k^{(r)}\lambda^k=
 (1-\lambda)^{-r}\exp\left(-\frac{z\lambda}{1-\lambda}\right), 
     \quad p_k^{(r)}=0\ \mbox{for}\ k<0, \\
\disp 
 q_k^{(r)}(z)=p_k^{(r)}(-z). 
\end{array}   \label{GF:P5}
\end{equation}
We define the polynomials $R_{m,n}=R_{m,n}^{(r)}(z)$ by 
\begin{equation}
R_{m,n}^{(r)}(z)=
 \left|
  \begin{array}{cccccccc}
   q_1^{(r)}        & q_0^{(r)}        & \cdots           & q_{-m+2}^{(r)}   &
   q_{-m+1}^{(r)}   & \cdots           & q_{-m-n+3}^{(r)} & q_{-m-n+2}^{(r)} \\
   q_3^{(r)}        & q_2^{(r)}        & \cdots           & q_{-m+4}^{(r)}   &
   q_{-m+3}^{(r)}   & \cdots           & q_{-m-n+5}^{(r)} & q_{-m-n+4}^{(r)} \\
   \vdots           & \vdots           & \ddots           & \vdots           &
   \vdots           & \ddots           & \vdots           & \vdots           \\
   q_{2m-1}^{(r)}   & q_{2m-2}^{(r)}   & \cdots           & q_m^{(r)}        &
   q_{m-1}^{(r)}    & \cdots           & q_{m-n+1}^{(r)}  & q_{m-n}^{(r)}    \\
   p_{n-m}^{(r)}    & p_{n-m+1}^{(r)}  & \cdots           & p_{n-1}^{(r)}    &
   p_n^{(r)}        & \cdots           & p_{2n-2}^{(r)}   & p_{2n-1}^{(r)}   \\
   \vdots           & \vdots           & \ddots           & \vdots           &
   \vdots           & \ddots           & \vdots           & \vdots           \\
   p_{-n-m+4}^{(r)} & p_{-n-m+5}^{(r)} & \cdots           & p_{-n+3}^{(r)}   &
   p_{-n+4}^{(r)}   & \cdots           & p_2^{(r)}        & p_3^{(r)}        \\
   p_{-n-m+2}^{(r)} & p_{-n-m+3}^{(r)} & \cdots           & p_{-n+1}^{(r)}   &
   p_{-n+2}^{(r)}   & \cdots           & p_0^{(r)}        & p_1^{(r)}
  \end{array}
 \right|,     \label{P5:rat:tau}
\end{equation}
for $m,n \in \BZ_{\ge 0}$ and by 
\begin{equation}
R_{m,n}=(-1)^{m(m+1)/2}R_{-m-1,n}, \quad 
R_{m,n}=(-1)^{n(n+1)/2}R_{m,-n-1}, 
\end{equation}
for $m,n \in \BZ_{<0}$, respectively. 
Then, setting 
\begin{equation}
R_{m,n}^{(r)}(z)=S_{m,n}(t,s), \label{P5:R-S}
\end{equation}
with 
\begin{equation}
z=\frac{t}{2}, \quad r=2s-m+n, \label{ztrs}
\end{equation}
we see that 
\begin{equation}
\begin{array}{l}
\medskip
\disp
q=-\frac{S_{m,n-1}(t,s)  S_{m-1,n}(t,s)}
        {S_{m-1,n}(t,s-1)S_{m,n-1}(t,s+1)}, \\
\disp
p=-\frac{2n-1}{4}
   \frac{S_{m-1,n}(t,s-1)S_{m,n-1}(t,s+1)S_{m-1,n-2}(t,s)}
        {S_{m-1,n-1}^2(t,s)S_{m,n-1}(t,s)}, 
\end{array}   \label{sol:P5}
\end{equation}
give a family of rational solutions to the Hamilton system S$_{\rm V}$
 for the parameters 
\begin{equation}
\kappa_{\infty}=s, \quad \kappa_0=s-m+n, \quad \theta=m+n-1. \label{para:P5}
\end{equation}
\end{prop}

Let us consider the degeneration of the algebraic solutions of S$_{\rm VI}$. 
Applying the B\"acklund transformation $s_0$ to the solutions in Theorem
\ref{detV}, we obtain the following corollary. 

\begin{coro}\label{toV}
Let $S_{m,n}=S_{m,n}(x;a,b)$ be polynomials given 
in Theorem \ref{detV}. 
Then, for $m,n \in \BZ$, 
\begin{equation}
q=x\frac{S_{m,n-1}^{(1,0)} S_{m-1,n}^{(1,0)}}
        {S_{m-1,n}^{(1,-1)}S_{m,n-1}^{(1,1)}}, \quad 
p=\frac{2n-1}{2x(1-x)}
  \frac{S_{m-1,n}^{(1,-1)} S_{m,n-1}^{(1,1)}S_{m-1,n-2}^{(2,0)}}
       {S_{m-1,n-1}^{(1,0)}S_{m,n-1}^{(1,0)}S_{m-1,n-1}^{(2,0)}}, \label{sol'}
\end{equation}
where we denote as $S_{m,n}^{(k,l)}=S_{m,n}(x;a+k,b+l)$, 
satisfy S$_{\rm VI}$ for the parameters 
\begin{equation}
\kappa_{\infty}=b, \quad \kappa_0=b-m+n, \quad \kappa_1=a+m+n, \quad \theta=-a, 
\label{para'}
\end{equation}
with $x^2=t$. 
\end{coro}

It is easy to see that by putting 
\begin{equation}
t \to 1-\ep t, \quad a=\ep^{-1}, 
\end{equation}
S$_{\rm VI}$ with (\ref{para'}) is reduced to S$_{\rm V}$ with
(\ref{para:P5}) in the limit of $\ep \to 0$. 

Next, we investigate the degeneration of $R_{m,n}^{(i,j)}$ given in
Definition \ref{def:V}. 
Putting 
\begin{equation}
x \to -(1-\ep t)^{\frac{1}{2}}, \quad c=\ep^{-1}+s+n-\frac{1}{2}, \quad d=2s-m+n, 
\end{equation}
we see that the generating function (\ref{GF:V}) degenerates as 
\begin{eqnarray}
G&=&(1-\lambda)^{-d}
    \exp\left\{c\left[\log(1-\lambda)-\log(1+x^{\pm 1}\lambda)\right]\right\} \cr
 &=&(1-\lambda)^{-r}\exp\left(\mp \frac{z\lambda}{1-\lambda}+O(\ep)\right), 
\end{eqnarray}
where we use (\ref{ztrs}). 
Then, we have 
\begin{equation}
\lim_{\ep \to 0}p_k^{(c,d)}(x)=p_k^{(r)}(z), \quad 
\lim_{\ep \to 0}q_k^{(c,d)}(x)=q_k^{(r)}(z), 
\end{equation}
which gives 
\begin{equation}
\lim_{\ep \to 0}R_{m,n}^{(i,j)}(x)=R_{m,n}^{(r+j)}(z). 
\end{equation}

Finally, it is easy to see that (\ref{sol'}) yields to (\ref{sol:P5}). 

\begin{rem}
As we mentioned in Section \ref{intr}, 
Kirillov and Taneda have introduced ``generalized Umemura polynomials''
 for P$_{\rm VI}$ in the context of combinatorics and shown that these
 polynomials degenerate to $S_{m,n}=S_{m,n}(t,s)$ defined in Proposition
 \ref{S5:rat} in some limit~\cite{KT1,KT2,KT3}. 
\end{rem}

\begin{rem}
The polynomials $p_k^{(r)}$ (and $q_k^{(r)}$) defined by (\ref{GF:P5}) 
are essentially the Laguerre polynomials, namely, 
$p_k^{(r)}(z)=L_k^{(r-1)}(z)$. 
The above degeneration corresponds to that from the Jacobi polynomials
 to the Laguerre polynomials. 
\end{rem}

\subsection{Degeneration from P$_{\rm VI}$ to P$_{\rm III}$}
Next, we consider P$_{\rm III}$ 
\begin{equation}
\frac{d^2y}{dt^2}=
\frac{1}{y}\left(\frac{dy}{dt}\right)^2-\frac{1}{t}\frac{dy}{dt}
-\frac{4}{t}\left[\eta_{\infty}\theta_{\infty}y^2+\eta_0(\theta_0+1)\right]
+4\eta_{\infty}^2 y^3-\frac{4\eta_0^2}{y}, \label{P3}
\end{equation}
which is equivalent to the Hamilton system~\cite{O4} 
\begin{equation}
\hskip-40pt
\mbox{S$_{\rm III}$~:} \hskip30pt
q'=\frac{\partial H}{\partial p}, \quad p'=-\frac{\partial H}{\partial q}, \quad
'=t\frac{d}{dt}, \label{cano3}
\end{equation}
with the Hamiltonian 
\begin{equation}
H=2q^2p^2-[2\eta_{\infty}tq^2+(2\theta_0+1)q+2\eta_0 t]p
         +\eta_{\infty}(\theta_{\infty}+\theta_0)tq.     \label{H3}
\end{equation}
This system can be also derived from S$_{\rm VI}$, directly, by degeneration. 
This process is achieved by putting 
\begin{equation}
t \to \ep^2 t^2, \quad q \to \ep tq, \quad p \to \ep^{-1}t^{-1}p, \label{6to3:var}
\end{equation}
\begin{equation}
\kappa_{\infty} \to \eta_{\infty}\ep^{-1}+\theta_{\infty}^{(1)}, \quad 
\kappa_0 \to -\eta_0 \ep^{-1}+\theta_0^{(1)}+1, \quad 
\kappa_1 \to -\eta_{\infty}\ep^{-1}+\theta_{\infty}^{(2)}, \quad 
\theta \to \eta_0 \ep^{-1}+\theta_0^{(2)},  \label{6to3:para}
\end{equation}
\begin{equation}
H \to -\frac{1}{2}(H+qp), \label{6to3:H}
\end{equation}
and taking the limit of $\ep \to 0$. 
In fact, the system (\ref{cano6}) with the Hamiltonian (\ref{H6}) is
reduced to (\ref{cano3}) with (\ref{H3}) by this procedure, 
where we put 
\begin{equation}
\theta_{\infty}=\theta_{\infty}^{(1)}+\theta_{\infty}^{(2)}, \quad 
\theta_0=\theta_0^{(1)}+\theta_0^{(2)}. 
\end{equation}

It is known that the rational solutions of S$_{\rm III}$ are expressed
as follows~\cite{P3:rat}. 

\begin{prop}\label{S3:rat}
Let $p_k=p_k^{(r)}(t),~k \in \BZ$, be polynomials defined by 
\begin{equation}
\sum_{k=0}^{\infty}p_k^{(r)}\lambda^k=
(1+\lambda)^r \exp\left(-{t\lambda}\right), 
\quad p_k^{(r)}=0\ \mbox{for}\ k<0.  \label{GF:P3}
\end{equation}
We define a family of polynomials $R_n^{(r)}=R_n^{(r)}(t)$ by 
\begin{equation}
R_n^{(r)}(t)=
 \left|
  \begin{array}{cccc}
   p_n^{(r)}      & \cdots & p_{2n-2}^{(r)} & p_{2n-1}^{(r)}   \\
   \vdots         & \ddots & \vdots         & \vdots           \\
   p_{-n+4}^{(r)} & \cdots & p_2^{(r)}      & p_3^{(r)}        \\
   p_{-n+2}^{(r)} & \cdots & p_0^{(r)}      & p_1^{(r)}
  \end{array}
 \right|,     \label{P3:rat:tau}
\end{equation}
for $n \in \BZ_{\ge 0}$ and by 
\begin{equation}
R_n=(-1)^{n(n+1)/2}R_{-n-1}, 
\end{equation}
for $n \in \BZ_{<0}$, respectively. 
Then, 
\begin{equation}
q=\frac{R_{n-1}^{(r+1)}R_n^{(r)}}
       {R_n^{(r+1)}R_{n-1}^{(r)}}, \quad 
p=-\frac{2n-1}{2}
   \frac{R_n^{(r+1)}R_{n-2}^{(r+1)}}
        {R_{n-1}^{(r+1)}R_{n-1}^{(r+1)}}, \label{sol:P3}
\end{equation}
give the rational solutions of S$_{\rm III}$ for the parameters 
\begin{equation}
\theta_{\infty}=r+\frac{1}{2}+n, \quad \theta_0+1=-r-\frac{1}{2}+n, \label{para:P3}
\end{equation}
with 
\begin{equation}
\eta_{\infty}=\eta_0=\frac{1}{2}. \label{scale-para}
\end{equation}
\end{prop}

Before discussing the degeneration to the rational solutions of 
S$_{\rm III}$, 
we slightly rewrite the determinant expression in Definition \ref{def:V}
for convenience. 

\begin{lem}\label{def:III}
Let $\bar{p}_k=\bar{p}_k^{(\bar{c},\bar{d})}(x)$ and 
$\bar{q}_k=\bar{q}_k^{(\bar{c},\bar{d})}(x)$, 
$k \in \BZ$, be two sets of polynomials defined by 
\begin{equation}
\begin{array}{c}
\disp 
\sum_{k=0}^{\infty}\bar{p}_k^{(\bar{c},\bar{d})}(x)\lambda^k=
\bar{G}(x;\bar{c},\bar{d};\lambda), \quad 
\bar{p}_k^{(\bar{c},\bar{d})}(x)=0\ \mbox{for}\ k<0, \\
\disp 
\bar{q}_k^{(\bar{c},\bar{d})}(x)=\bar{p}_k^{(\bar{c},\bar{d})}(x^{-1}), 
\end{array}   \label{def:pq:III}
\end{equation}
respectively, 
where the generating function $\bar{G}(x;\bar{c},\bar{d};\lambda)$ is given by 
\begin{equation}
\bar{G}(x;\bar{c},\bar{d};\lambda)=
(1-\lambda)^{\bar{d}-1}(1+x\lambda)^{-\bar{c}}. \label{GF:III}
\end{equation}
Define $\bar{R}_{m,n}=\bar{R}_{m,n}(x;\bar{c},\bar{d})$ in terms of the 
same determinant as (\ref{P6:alg:tau}) with entries $p_k$ and $q_k$ 
replaced by $\bar{p}_k$ and $\bar{q}_k$, respectively. 
Then, we have 
\begin{equation}
\bar{R}_{m,n}(x;\bar{c},\bar{d})=S_{m,n}(x;a,b), \label{S-R:III}
\end{equation}
with 
\begin{equation}
\bar{c}=a+b+n-\frac{1}{2}, \quad \bar{d}=a-b+m+\frac{1}{2}. \label{abcd:III}
\end{equation}
\end{lem}

\begin{rem}
The polynomials $\bar{p}_k$ and $\bar{q}_k$ are also expressed by the 
Jacobi polynomials as 
\begin{equation}
\bar{p}_k^{(\bar{c},\bar{d})}(x)=(-1)^kP_k^{(\bar{d}-1-k,\bar{c}-\bar{d})}(1+2x). 
\end{equation}
\end{rem}

Let us consider the degeneration of the algebraic solutions of S$_{\rm VI}$. 
Applying the B\"acklund transformation $s_1s_0$ to the solutions in
Theorem \ref{detV}, we obtain the following corollary. 

\begin{coro}\label{toIII}
Let $\bar{R}_{m,n}=\bar{R}_{m,n}(x;\bar{c},\bar{d})$ be polynomials 
defined in Lemma \ref{def:III}. 
Then, for $m,n \in \BZ$, 
\begin{equation}
q=x\frac{\bar{R}_{m,n-1}^{(0,1)}\bar{R}_{m-1,n}^{(1,0)}}
        {\bar{R}_{m-1,n}^{(0,1)}\bar{R}_{m,n-1}^{(1,0)}}, \quad 
p=\frac{2n-1}{2x(1-x)}
  \frac{\bar{R}_{m-1,n}^{(0,1)}\bar{R}_{m,n-1}^{(1,0)}\bar{R}_{m-1,n-2}^{(0,1)}}
       {\bar{R}_{m-1,n-1}^{(0,0)}\bar{R}_{m,n-1}^{(0,1)}\bar{R}_{m-1,n-1}^{(1,1)}}, 
\label{sol''}
\end{equation}
where we denote as 
$\bar{R}_{m,n}^{(i,j)}=\bar{R}_{m,n}(x;\bar{c}+i,\bar{d}+j)$, 
satisfy S$_{\rm VI}$ for the parameters 
\begin{equation}
\kappa_{\infty}=-b, \quad \kappa_0=b-m+n, \quad \kappa_1=a+m+n, \quad \theta=-a, 
\label{para''}
\end{equation}
under the setting of (\ref{S-R:III}),(\ref{abcd:III}) and $x^2=t$. 
\end{coro}

According to (\ref{6to3:para}) and (\ref{para''}), we fix as
(\ref{scale-para}) and put 
\begin{equation}
a=\frac{1}{2}\left(-\ep^{-1}+r+\frac{1}{2}-m+\zeta\right), \quad 
b=\frac{1}{2}\left(-\ep^{-1}-r-\frac{1}{2}+m+\zeta\right), 
\end{equation}
where $\zeta$ is a quantity of $O(1)$. 
Then, we have 
\begin{equation}
\begin{array}{c}
\smallskip
\disp 
\theta_{\infty}^{(1)}=\frac{1}{2}\left(r+\frac{1}{2}-m-\zeta\right), \quad 
\theta_{\infty}^{(2)}=\frac{1}{2}\left(r+\frac{1}{2}+m+\zeta\right)+n, \\
\disp 
\theta_0^{(1)}+1=-\frac{1}{2}\left(r+\frac{1}{2}+m-\zeta\right)+n, \quad 
\theta_0^{(2)}  =-\frac{1}{2}\left(r+\frac{1}{2}-m+\zeta\right). 
\end{array}
\end{equation}
Setting as (\ref{6to3:var}) and (\ref{6to3:H}), 
we see that S$_{\rm VI}$ with (\ref{para''}) is reduced to S$_{\rm III}$
with (\ref{para:P3}) in the limit of $\ep \to 0$. 
Note that $m$ vanishes in (\ref{para:P3}). 
Then, it is possible to put $m=0$ without losing generality in this
limiting procedure. 

Next, we investigate the degeneration of
$\bar{R}_n^{(i,j)}=\bar{R}_{-1,n}^{(i,j)}=\bar{R}_{0,n}^{(i,j)}$. 
Putting 
\begin{equation}
x \to \ep t, \quad 
\bar{c}=-\ep^{-1}+\zeta+n-\frac{1}{2}, \quad \bar{d}=r+1, 
\end{equation}
we find that the generating function (\ref{GF:III}) degenerates as 
\begin{equation}
\bar{G}=(1-\lambda)^{\bar{d}-1}\exp\left[-\bar{c}\log(1+x\lambda)\right]
       =(1-\lambda)^r\exp\left[t\lambda+O(\ep)\right]. 
\end{equation}
Then, we have 
\begin{equation}
\lim_{\ep \to 0}\bar{p}_k^{(\bar{c},\bar{d})}(x)=(-1)^kp_k^{(r)}(t), 
\end{equation}
which gives 
\begin{equation}
\lim_{\ep \to 0}\bar{R}_n^{(i,j)}(x)=(-1)^{n(n+1)/2}R_n^{(r+j)}(t). 
\end{equation}

Finally, it is easy to see that (\ref{sol''}) is lead to (\ref{sol:P3})
in the above limit. 

\begin{rem}
The polynomials $p_k^{(r)}$ defined by (\ref{GF:P3}) are also the 
Laguerre polynomials, namely, $p_k^{(r)}(t)=L_k^{(r-k)}(t)$. 
Then, the above degeneration also corresponds to that from the Jacobi
 polynomials to the Laguerre polynomials. 
\end{rem}

Similarly, the rational solutions of P$_{\rm V}$ and P$_{\rm III}$ given
in Proposition \ref{S5:rat} and \ref{S3:rat}, respectively, degenerate to
those of P$_{\rm II}$. 
We give detail in Appendix \ref{coal:to2}. 
Therefore, the coalescence cascade (\ref{coal:alg}) is obtained.

\section{Relationship to the original Umemura polynomials \label{rel:Ume}}
In this section, we show that the original Umemura polynomials for
P$_{\rm VI}$ are understood as a special case of our polynomials
$V_{m,n}(x;a,b)$ introduced in Section \ref{constr}.

\subsection{Umemura polynomials associated with P$_{\rm VI}$}
First, we briefly summarize the derivation of the original Umemura
polynomials for P$_{\rm VI}$~\cite{Um}. 
Set the parameters $b_i~(i=1,2,3,4)$ as 
\begin{equation}
b_1=\frac{1}{2}(\kappa_0+\kappa_1), \quad b_2=\frac{1}{2}(\kappa_0-\kappa_1), \quad 
b_3=\frac{1}{2}(\theta-1+\kappa_{\infty}), \quad b_4=\frac{1}{2}(\theta-1-\kappa_{\infty}), 
\end{equation}
namely, 
\begin{equation}
b_1=\frac{1}{2}(\aa_4+\aa_3), \quad b_2=\frac{1}{2}(\aa_4-\aa_3), \quad 
b_3=\frac{1}{2}(\aa_0-1+\aa_1), \quad b_4=\frac{1}{2}(\aa_0-1-\aa_1).  \label{b:aa}
\end{equation}
Umemura has shown that 
\begin{equation}
q=\frac{(\aa+\beta)^2t\pm(\aa^2-\beta^2)\sqrt{t(t-1)}}{(\aa-\beta)^2+4\aa \beta t}, \quad 
p=\frac{\aa q-(\aa+\beta)/2}{q(q-1)}, \label{seed:Ume}
\end{equation}
give an algebraic solution of the Hamilton system S$_{\rm VI}$ for the parameters 
\begin{equation}
(b_1,b_2,b_3,b_4)=\left(\aa,\beta,-\frac{1}{2},0 \right). \label{seed:Ume:para}
\end{equation}
Substituting the solution of upper sign into the Hamiltonian (\ref{H6}),
one obtain 
\begin{equation}
H=\frac{1}{4}
  \left[-(\aa+\beta)+(\aa+\beta)^2+2\aa t-2(\aa^2+\beta^2)t
        +2(\aa^2-\beta^2)\sqrt{t(t-1)}\right]. 
\end{equation}
Application of the translation 
\begin{equation}
(b_1,b_2,b_3,b_4) \to (b_1,b_2,b_3,b_4)+n(0,0,1,0), \quad n \in \BZ, 
\end{equation}
to the seed solution (\ref{seed:Ume}) with (\ref{seed:Ume:para})
generates a sequence of algebraic solutions $(q_n,p_n)$. 
Let $\tau_n$ be a $\tau$-function with respect to the solution $(q_n,p_n)$. 
Okamoto has pointed out that $\tau_n$ satisfy the Toda equation~\cite{O1} 
\begin{equation}
\frac{\tau_{n+1}\tau_{n-1}}{\tau_n^2}=
\frac{d}{dt}(\log \tau_n)'+(b_1+b_3+n)(b_3+b_4+n), \quad '=t(t-1)\frac{d}{dt}. 
\label{Toda:Okamoto}
\end{equation}
Define a family of functions $T_n$ for $n \in \BZ$ by 
\begin{equation}
(\log \tau_n)'=\left(\log T_n \right)'+H-n\left(\aa t-\frac{\aa+\beta}{2}\right). 
\end{equation}
Then, Toda equation (\ref{Toda:Okamoto}) yields to 
\begin{equation}
\begin{array}{l}
\medskip
\disp
 T_{n+1}T_{n-1}=
  t(t-1)\left[\frac{d^2 T_n}{dt^2}T_n-\left(\frac{d T_n}{dt}\right)^2\right]
                  +(2t-1)\frac{d T_n}{dt}T_n \\
\disp
 \hskip70pt
   +\left\{\frac{1}{4}\left[-2(\aa^2+\beta^2)
                            +(\aa^2-\beta^2)\frac{2t-1}{\sqrt{t(t-1)}}\right]
         +\left(n-\frac{1}{2}\right)^2\right\}T_n^2. 
\end{array}   \label{Toda:T}
\end{equation}
Moreover, introducing a new variable $v$ as 
\begin{equation}
v=\sqrt{\frac{t}{t-1}}+\sqrt{\frac{t-1}{t}},  
\end{equation}
one find that $T_n$ are generated by the recurrence relation 
\begin{equation}
\begin{array}{l}
\medskip
\disp
 T_{n+1}T_{n-1}=
    \frac{1}{4}(v^2-4)\left[(v^2-4)\frac{d^2 T_n}{dv^2}+v\frac{d T_n}{dv}\right]T_n
    -\frac{1}{4}(v^2-4)^2 \left(\frac{d T_n}{dv}\right)^2 \\
\disp
 \hskip70pt +\left\{\frac{1}{4}\left[-2(\aa^2+\beta^2)+(\aa^2-\beta^2)v\right]
                        +\left(n-\frac{1}{2}\right)^2\right\}T_n^2, 
\end{array}
\end{equation}
with the initial conditions $T_0=T_1=1$. 
It is shown that $T_n$ for $n \in \BZ_{\ge 0}$ are polynomials in
$\aa,\beta$ and $v$, and $\deg_v T_n=n(n-1)/2$. 
These polynomials are called Umemura polynomials associated with 
P$_{\rm VI}$.

\subsection{Correspondence of the seed solution}
We investigate how the Umemura's seed solution (\ref{seed:Ume}) with
(\ref{seed:Ume:para}) is related to ours, 
\begin{equation}
q=f_4=x, \quad p=f_2=\frac{1}{2}\left(a+b-\frac{1}{2}\right)x^{-1}, \label{seed:f}
\end{equation}
with
\begin{equation}
(\aa_0,\aa_1,\aa_2,\aa_3,\aa_4)=\left(a,b,\frac{1}{2}-a-b,a,b \right). \label{seed:para}
\end{equation}
In addition to the B\"acklund transformations stated in Section
\ref{sym}, 
it is known that P$_{\rm VI}$ admits the outer symmetry as follows~\cite{O1}, 
\begin{equation}
\begin{array}{|c||c|c|c|c|}
\hline
            & \aa_0 \quad \aa_1 \quad \aa_2 \quad \aa_3 \quad \aa_4 
            & t & f_4 & f_2  \\
\hline
\sigma_{01} & \aa_1 \quad \aa_0 \quad \aa_2 \quad \aa_3 \quad \aa_4 
            & 1-t 
            & \disp \frac{(1-t)f_4}{f_0} 
            & \disp \frac{f_0(f_0f_2+\aa_2)}{t(t-1)}  \\
\sigma_{03} & \aa_3 \quad \aa_1 \quad \aa_2 \quad \aa_0 \quad \aa_4 
            & \disp \frac{1}{t} 
            & \disp \frac{f_4}{t}
            & tf_2  \\
\sigma_{04} & \aa_4 \quad \aa_1 \quad \aa_2 \quad \aa_3 \quad \aa_0 
            & \disp \frac{t}{t-1} 
            & \disp \frac{f_0}{1-t}
            & (1-t)f_2  \\
\sigma_{13} & \aa_0 \quad \aa_3 \quad \aa_2 \quad \aa_1 \quad \aa_4 
            & \disp \frac{t}{t-1} 
            & \disp \frac{f_4}{f_3}
            & -f_3(f_3f_2+\aa_2)  \\
\sigma_{14} & \aa_0 \quad \aa_4 \quad \aa_2 \quad \aa_3 \quad \aa_1 
            & \disp \frac{1}{t}   
            & \disp \frac{1}{f_4}
            & -f_4(f_4f_2+\aa_2)  \\
\sigma_{34} & \aa_0 \quad \aa_1 \quad \aa_2 \quad \aa_4 \quad \aa_3 
            & 1-t & -f_3 & -f_2  \\
\hline
\end{array}~.   \label{outer}
\end{equation}

\begin{prop}
Umemura's seed solution (\ref{seed:Ume}) with (\ref{seed:Ume:para}) is
 obtained by applying the B\"acklund transformation defined by 
\begin{equation}
\sigma=\sigma_{13}s_3s_2s_1, 
\end{equation}
to ours (\ref{seed:f}) with (\ref{seed:para}), where we put 
\begin{equation}
\aa=\frac{1}{2}-a, \quad \beta=b. \label{corre:para}
\end{equation}
\end{prop}
\noindent
{\it Proof.}\quad 
First, we check for the parameters. 
Application of $\sigma$ to (\ref{seed:para}) gives 
\begin{equation}
(\aa_0,\aa_1,\aa_2,\aa_3,\aa_4)=
\left(\frac{1}{2},-\frac{1}{2},a,\frac{1}{2}-a-b,\frac{1}{2}-a+b\right), 
\end{equation}
which coincides with (\ref{seed:Ume:para}) by using (\ref{b:aa}) and (\ref{corre:para}). 

Next, we verify the correspondence of $q=f_4$. 
We have 
\begin{equation}
\sigma(f_4)
=\frac{f_2f_4+\aa_1+\aa_2}{f_2f_3+\aa_1+\aa_2}
=\frac{\disp \frac{1}{2}-a+b}
      {\disp \left(\frac{1}{2}-a+b\right)+\left(\frac{1}{2}-a-b\right)x^{-1}}. 
\label{seed:Ume'}
\end{equation}
Note that $x$ is now given by 
\begin{equation}
x=\mp \sqrt{\frac{t}{t-1}}, 
\end{equation}
due to the action of $\sigma_{13}$. 
Thus, the expression (\ref{seed:Ume'}) is equivalent to the first of
(\ref{seed:Ume}). 
It is possible to check for $p=f_2$ in similar way. 
\hfill \qed

\subsection{Relationship to the original Umemura polynomials}
The above discussion on the seed solution suggests that the family of
polynomials $V_{m,n}(x;a,b)$ constructed in Section \ref{constr}
corresponds to the original Umemura polynomials under the setting of
(\ref{corre:para}) and 
\begin{equation}
x=-\sqrt{\frac{t}{t-1}}. \label{xt}
\end{equation}
Notice that, from (\ref{Toda:T}) and (\ref{xt}), 
$T_n=T_n(x;\aa,\beta)$ satisfy the recurrence relation 
\begin{equation}
4T_{n+1}T_{n-1}=x^{-1}\left[(x^2-1)^2{\cal D}^2 
                -\aa^2(x+1)^2+\beta^2(x-1)^2+(2n-1)^2x \right]T_n \cdot T_n,  \label{Toda:T:x}
\end{equation}
with $T_0=T_1=1$. 

\begin{thm}\label{TbyV}
We have 
\begin{equation}
T_n(x;\aa,\beta)=2^{-2n(n-1)}(-x)^{-n(n-1)/2}V_{-n,-n}(x;a+n,b), \label{T-V}
\end{equation}
with (\ref{corre:para}). 
\end{thm}

We prove Theorem \ref{TbyV} by showing that both hand sides of
(\ref{T-V}) satisfy the same recurrence relation and initial
conditions. 
Let $\widehat{T}_{30}$ be the translation operator defined by 
\begin{equation}
\widehat{T}_{30}=T_{34}\widehat{T}_{34}T_{03}^{-1}. 
\end{equation}
Then, we have a Toda equation 
\begin{equation}
\widehat{T}_{30}(\tau_0)\widehat{T}_{30}^{-1}(\tau_0)=t^{-\frac{1}{2}}
\left[(t-1)\frac{d}{dt}(\log \tau_0)'-(\log \tau_0)'
      +\frac{1}{4}(\aa_0-\aa_3)(\aa_0-\aa_3-2)+\frac{3}{4}\right]\tau_0^2. 
\end{equation}
For simplicity, we denote as 
\begin{equation}
\bar{\tau}_n=\widehat{T}_{30}^n(\tau_0), \quad n \in \BZ, 
\end{equation}
namely $\bar{\tau}_n=\tau_{-n,0,n,n}$ in the notation of (\ref{tau}). 
The above Toda equation is expressed as 
\begin{equation}
\bar{\tau}_{n+1}\bar{\tau}_{n-1}=t^{-\frac{1}{2}}
\left[(t-1)\frac{d}{dt}(\log \bar{\tau}_n)'-(\log \bar{\tau}_n)'
      +\frac{(\aa_0-\aa_3-2n)(\aa_0-\aa_3-2n-2)}{4}+\frac{3}{4}\right]\bar{\tau}_n^2. 
\label{Toda:tau:bar}
\end{equation}
In the following, we restrict our discussion to the algebraic solutions. 
According to (\ref{tau-sig}) and (\ref{sig-V}), 
we introduce $\bar{V}_n=\bar{V}_n(x;a,b)$ as 
\begin{equation}
\bar{\tau}_n=\omega_n \bar{V}_n 
        (x-1)^{(a-\frac{1}{2})^2+\frac{1}{2}}
            x^{-(a-\frac{1}{2})^2-b^2-n(n+1)-\frac{1}{8}}
        (x+1)^{b^2+\frac{1}{2}}, \label{tau-sig:bar}
\end{equation}
where $\omega_n=\omega_{-n,0,n,n}$. 
Substituting (\ref{tau-sig:bar}) and $\aa_0=\aa_3=a$ into Toda equation
(\ref{Toda:tau:bar}) and noticing 
\begin{equation}
\omega_{n+1}\omega_{n-1}=-\frac{1}{16}\omega_n^2, 
\end{equation}
we find that $\bar{V}_n=\bar{V}_n(x;a,b)$ are generated by the
recurrence relation 
\begin{equation}
-\frac{1}{4}\bar{V}_{n+1}\bar{V}_{n-1}=
\left[(x^2-1)^2{\cal D}^2
       -\left(a-\frac{1}{2}\right)^2(x+1)^2+b^2(x-1)^2+(2n+1)^2x\right]
\bar{V}_n \cdot \bar{V}_n, 
\end{equation}
with the initial conditions $\bar{V}_{-1}=\bar{V}_0=1$. 
By construction, it is easy to see that 
\begin{equation}
\bar{V}_n(x;a,b)=V_{n,n}(x;a-n,b).  
\end{equation}
Moreover, we introduce $\bar{T}_n=\bar{T}_n(x;a,b)$ as 
\begin{equation}
\bar{T}_n=2^{-2n(n-1)}(-x)^{-n(n-1)/2}\bar{V}_{-n}. 
\label{T-V:bar}
\end{equation}
Then, $\bar{T}_n$ satisfy the recurrence relation 
\begin{equation}
4\bar{T}_{n+1}\bar{T}_{n-1}=
x^{-1}\left[(x^2-1)^2{\cal D}^2
       -\left(a-\frac{1}{2}\right)^2(x+1)^2+b^2(x-1)^2+(2n-1)^2x\right]
\bar{T}_n \cdot \bar{T}_n, \label{Toda:T:bar}
\end{equation}
with $\bar{T}_0=\bar{T}_1=1$. 

Comparing (\ref{Toda:T:x}) with (\ref{Toda:T:bar}), we find 
\begin{equation}
T_n(x;\aa,\beta)=\bar{T}_n(x;a,b), \label{corre:tau}
\end{equation}
under the setting of (\ref{corre:para}), 
which is nothing but Theorem \ref{TbyV}. 

\begin{rem}
Toda equation (\ref{Toda:Okamoto}) can be regarded as the recurrence
relation with respect to the translation operator 
\begin{equation}
T_{01}=T_{34}^{-1}T_{14}T_{03}, 
\end{equation}
which acts on the parameters as 
\begin{equation}
T_{01}(\aa_0,\aa_1,\aa_2,\aa_3,\aa_4)=
(\aa_0,\aa_1,\aa_2,\aa_3,\aa_4)+(1,1,-1,0,0). 
\end{equation}
Theorem \ref{TbyV}, namely (\ref{corre:tau}), is consistent with the relation 
\begin{equation}
T_{01}\sigma=\sigma \widehat{T}_{30}^{-1}. 
\end{equation}
\end{rem}

From the discussion of the previous sections and (\ref{T-V:bar}), 
it is clear that $T_n$ for $n \in \BZ_{\ge 0}$ are polynomials in
$\aa,\beta$ and $v$, and $\deg_v T_n=n(n-1)/2$ under the setting of 
\begin{equation}
v=-\left(x+x^{-1}\right). 
\end{equation}

\medskip

\noindent 
{\bf Acknowledgment}\quad 
The author would like to thank Prof. M. Noumi, Prof. Y. Yamada and
Prof. K. Kajiwara for useful suggestions and discussions. 
Especially, he owes the initial steps of this work to discussions with
them. 
The author would also express his sincere thanks to Prof. Kirillov and 
Dr. Taneda for stimulating discussions.

\appendix
\section{Degeneration of rational solutions \label{coal:to2}}
In this section, we show that the rational solutions of P$_{\rm V}$
and P$_{\rm III}$ degenerate to those of P$_{\rm II}$, 
\begin{equation}
\frac{d^2y}{dt^2}=2y^3-4ty+4\left(\aa+\frac{1}{2}\right).  \label{P2}
\end{equation}
As is known~\cite{O3}, 
P$_{\rm II}$ (\ref{P2}) is equivalent to the Hamilton system 
\begin{equation}
\hskip-40pt
\mbox{S$_{\rm II}$~:} \hskip30pt
q'=\frac{\partial H}{\partial p}, \quad p'=-\frac{\partial H}{\partial q}, \quad
'=\frac{d}{dt}, \label{cano2}
\end{equation}
with the Hamiltonian 
\begin{equation}
H=-2p^2-(q^2-2t)p+\aa q. \label{H2}
\end{equation}

The rational solutions of S$_{\rm II}$ are expressed as follows~\cite{P2:rat}. 
\begin{prop}\label{S2:rat}
Let $q_k=q_k(t),~k \in \BZ$, be polynomials defined by 
\begin{equation}
\sum_{k=0}^{\infty}q_k\lambda^k=\exp \left( t\lambda+\frac{\lambda^3}{3}\right), 
\quad q_k=0\ \mbox{for}\ k<0.  \label{GF:P2}
\end{equation}
We define $R_n=R_n(t)$ by 
\begin{equation}
R_n=
 \left|
  \begin{array}{cccc}
   q_n      & \cdots & q_{2n-2} & q_{2n-1} \\
   \vdots   & \ddots & \vdots   & \vdots   \\
   q_{-n+4} & \cdots & q_2      & q_3      \\
   q_{-n+2} & \cdots & q_0      & q_1
  \end{array}
 \right|,     \label{P2:rat:tau}
\end{equation}
for $n \in \BZ_{\ge 0}$ and by 
\begin{equation}
R_n=(-1)^{n(n+1)/2}R_{-n-1}, 
\end{equation}
for $n \in \BZ_{<0}$, respectively. 
Then, 
\begin{equation}
q=\frac{d}{dt}\log \frac{R_n}{R_{n-1}}, \quad 
p=\frac{2n-1}{2}\frac{R_nR_{n-2}}{R_{n-1}^2}, \label{sol:P2}
\end{equation}
give the rational solutions of S$_{\rm II}$ for the parameters 
\begin{equation}
\aa=n-\frac{1}{2}. \label{para:P2}
\end{equation}
\end{prop}

\subsection{From P$_{\rm V}$ to P$_{\rm II}$}
It is possible to derive the Hamilton system S$_{\rm II}$ 
from S$_{\rm V}$, directly, by degeneration. 
Putting 
\begin{equation}
t \to \ep^{-3}(1+2\ep^2 t), \quad 
q \to -1+2\ep q, \quad p \to \frac{1}{2}\ep^{-1}p, \label{5to2:var}
\end{equation}
\begin{equation}
\kappa_{\infty} \to \frac{\sigma}{4}\ep^{-3}+\kappa_{\infty}^{(0)}, \quad 
\kappa_0        \to      \frac{1}{4}\ep^{-3}+\kappa_0^{(0)}, \quad 
\theta \to 2\theta^{(0)}, \label{5to2:para}
\end{equation}
\begin{equation}
H \to \frac{1}{2}\ep^{-2}H-\frac{1}{2}\ep^{-3}\aa, \quad 
\aa=\theta^{(0)}+\frac{\kappa_0^{(0)}-\sigma \kappa_{\infty}^{(0)}}{2}, \label{5to2:H}
\end{equation}
with $\sigma=\pm 1$ and taking the limit of $\ep \to 0$, 
we find that the system (\ref{cano5}) with the Hamiltonian (\ref{H5}) is
reduced to (\ref{cano2}) with (\ref{H2}). 

We show that the rational solutions of S$_{\rm V}$ given in Proposition 
\ref{S5:rat} degenerate to those of S$_{\rm II}$ in Proposition \ref{S2:rat}. 
According to (\ref{5to2:para}) and (\ref{para:P5}), we put $\sigma=1$ and 
\begin{equation}
s=\frac{1}{4}\ep^{-3}, \quad 
\kappa_{\infty}^{(0)}=0, \quad \kappa_0^{(0)}=-m+n, \quad \theta^{(0)}=\frac{m+n-1}{2}. 
\end{equation}
Then, after the replacement (\ref{5to2:var}) and (\ref{5to2:H}), 
we find that S$_{\rm V}$ with (\ref{para:P5}) is reduced to
S$_{\rm II}$ with (\ref{para:P2}) in the limit of $\ep \to 0$. 
Note that $m$ vanishes in (\ref{para:P2}). 
Then, it is possible to put $m=0$ without loss of generality in this
limiting procedure. 

Next, we investigate the degeneration of $R_n^{(r)}=R_{-1,n}^{(r)}=R_{0,n}^{(r)}$. 
It is obvious that we have the following lemma. 
\begin{lem}
Let $\bar{p}_k=\bar{p}_k^{(r)}(z),~k \in \BZ$, be polynomials defined by 
\begin{equation}
\sum_{k=0}^{\infty}\bar{p}_k^{(r)}\lambda^k=
\exp \left[ \sum_{j=1}^{\infty}
           \left(-z+\frac{r}{j}\right)\lambda^j+\frac{r}{2}\lambda^2\right], 
\quad \bar{p}_k^{(r)}=0\ \mbox{for}\ k<0. \label{GF':P5}
\end{equation}
Then, we have 
\begin{equation}
R_n^{(r)}(z)=
 \left|
  \begin{array}{cccc}
   \bar{p}_n^{(r)}      & \cdots & \bar{p}_{2n-2}^{(r)} & \bar{p}_{2n-1}^{(r)} \\
   \vdots               & \ddots & \vdots               & \vdots               \\
   \bar{p}_{-n+4}^{(r)} & \cdots & \bar{p}_2^{(r)}      & \bar{p}_3^{(r)}      \\
   \bar{p}_{-n+2}^{(r)} & \cdots & \bar{p}_0^{(r)}      & \bar{p}_1^{(r)}
  \end{array}
 \right|. 
\end{equation}
\end{lem}
Put 
\begin{equation}
\lambda \to -\ep \lambda, \quad \bar{q}_k^{(r)}=(-\ep)^k \bar{p}_k^{(r)}, 
\end{equation}
and 
\begin{equation}
z \to \frac{1}{2}\ep^{-3}\left(1+2\ep^2t\right), \quad r=\frac{1}{2}\ep^{-3}+n. 
\end{equation}
Then, (\ref{GF':P5}) yields to 
\begin{equation}
\sum_{k=0}^{\infty}\bar{q}_k^{(r+j)}\lambda^k=
\exp \left( t\lambda+\frac{\lambda^3}{3}\right)
     \left[1-\ep \left( j\lambda+n\lambda+t\lambda^2+\frac{3}{8}\lambda^4\right)
            +O(\ep^2)\right]. 
\end{equation}
By using (\ref{GF:P2}), we obtain 
\begin{equation}
\bar{q}_k^{(r+j)}=
q_k-\ep jq_{k-1}-\ep\left(nq_{k-1}+tq_{k-2}+\frac{3}{8}q_{k-4}\right)+O(\ep^2). 
\label{q:bar}
\end{equation}
Since it is easy to see that 
\begin{equation}
\frac{d q_k}{dt}=q_{k-1}, 
\end{equation}
we have, from (\ref{P2:rat:tau}), 
\begin{equation}
R_n^{(r+j)}=(-\ep)^{-n(n+1)/2}
\left[R_n-\ep j\frac{d R_n}{dt}-\ep Q_n+O(\ep^2) \right], 
\end{equation}
where $Q_n$ denotes the contribution from the third term of (\ref{q:bar}). 

Finally, we verify the degeneration of the variables $q$ and $p$. 
The above procedure gives 
\begin{equation}
\begin{array}{l}
\smallskip
\disp 
-\frac{R_{n-1}^{(r-1)}R_n^{(r+1)}}
      {R_n^{(r-1)}R_{n-1}^{(r+1)}}=
-1+2\ep\frac{d}{dt}\log\frac{R_n}{R_{n-1}}+O(\ep^2), \\
\disp 
-\frac{R_n^{(r-1)}R_{n-1}^{(r+1)}R_{n-2}^{(r-1)}}
      {R_{n-1}^{(r)}R_{n-1}^{(r)}R_{n-1}^{(r-1)}}=
\ep^{-1}\frac{R_nR_{n-2}}{R_{n-1}^2}+O(1). 
\end{array}
\end{equation}
Thus, from (\ref{5to2:var}), 
we get (\ref{sol:P2}) in the limit of $\ep \to 0$.

\subsection{From P$_{\rm III}$ to P$_{\rm II}$}
It is well known that the Hamilton system S$_{\rm II}$ is derived from
S$_{\rm III}$ by degeneration~\cite{GtoP}. 
This process is achieved by putting 
\begin{equation}
t \to -\ep^{-3}(1-\ep^2 t), \quad q \to 1+\ep q, \quad p \to \ep^{-1}p,  \label{3to2:var}
\end{equation}
\begin{equation}
\theta_{\infty} \to -\ep^{-3}+\theta_{\infty}^{(0)}, \quad 
\theta_0        \to  \ep^{-3}+\theta_0^{(0)},        \label{3to2:para}
\end{equation}
\begin{equation}
H \to -\ep^{-2}H-\ep^{-3}\aa, \quad 
\aa=\frac{\theta_{\infty}^{(0)}+\theta^{(0)}}{2}, \label{3to2:H}
\end{equation}
and taking the limit of $\ep \to 0$. 

We show that the rational solutions of S$_{\rm III}$ given in
Proposition \ref{S3:rat} degenerate to those of S$_{\rm II}$ in
Proposition \ref{S2:rat}. 
From (\ref{3to2:para}), we put 
\begin{equation}
r=-\ep^{-3}, \quad 
\theta_{\infty}^{(0)}=n+\frac{1}{2}, \quad \theta_0^{(0)}=n-\frac{3}{2}. 
\end{equation}
Then, after replacing as (\ref{3to2:var}) and (\ref{3to2:H}), 
we see that S$_{\rm III}$ with (\ref{para:P3}) is reduced to 
S$_{\rm II}$ with (\ref{para:P2}) in the limit of $\ep \to 0$. 

Next, we investigate the degeneration of $R_n^{(r)}$ defined by
(\ref{GF:P3}) and (\ref{P3:rat:tau}). 
It is obvious that we have the following lemma. 
\begin{lem}
Let $\bar{p}_k=\bar{p}^{(r)}_k(t),~k \in \BZ$, be polynomials defined by 
\begin{equation}
\sum_{k=0}^{\infty}\bar{p}_k^{(r)}\lambda^k=
\exp \left[ \sum_{j=1}^{\infty}\frac{(-1)^{j-1}r}{j}\lambda^j
            -t \lambda +\frac{r}{2}\lambda^2\right], 
\quad \bar{p}_k^{(r)}=0\ \mbox{for}\ k<0. \label{GF':P3}
\end{equation}
Then, we have 
\begin{equation}
R_n^{(r)}(t)=
 \left|
  \begin{array}{cccc}
   \bar{p}_n^{(r)}      & \cdots & \bar{p}_{2n-2}^{(r)} & \bar{p}_{2n-1}^{(r)} \\
   \vdots               & \ddots & \vdots               & \vdots               \\
   \bar{p}_{-n+4}^{(r)} & \cdots & \bar{p}_2^{(r)}      & \bar{p}_3^{(r)}      \\
   \bar{p}_{-n+2}^{(r)} & \cdots & \bar{p}_0^{(r)}      & \bar{p}_1^{(r)}
  \end{array}
 \right|. 
\end{equation}
\end{lem}
Put 
\begin{equation}
\lambda \to -\ep \lambda, \quad \bar{q}_k^{(r)}=(-\ep)^k \bar{p}_k^{(r)}, 
\end{equation}
and 
\begin{equation}
t \to -\ep^{-3}(1-\ep^2t), \quad r=-\ep^{-3}. 
\end{equation}
Then, (\ref{GF':P3}) is written as 
\begin{equation}
\sum_{k=0}^{\infty}\bar{q}_k^{(r+j)}\lambda^k=
\exp \left( t\lambda+\frac{\lambda^3}{3}\right)
     \left[1+\ep \left( -j\lambda+\frac{1}{4}\lambda^4\right)
            +O(\ep^2)\right]. 
\end{equation}
By using (\ref{GF:P2}), we obtain 
\begin{equation}
\bar{q}_k^{(r+j)}=q_k+\ep \left(-j q_{k-1}+\frac{1}{4}q_{k-4} \right)+O(\ep^2). 
\label{q:bar:III}
\end{equation}
Thus, we have 
\begin{equation}
R_n^{(r)}=(-\ep)^{-n(n+1)/2}
\left[R_n+\ep \left( -j\frac{d R_n}{dt}+Q_n \right)+O(\ep^2) \right], 
\end{equation}
where $Q_n$ denotes the contribution from the term of $q_{k-4}$ in
(\ref{q:bar:III}). 

Finally, it is easy to see that (\ref{sol:P3}) is reduced to
(\ref{sol:P2}) by the above limiting procedures.

\section{Proof of Lemma \ref{V1}-\ref{V5} \label{PoL}}
We first note that the following contiguity relations hold by
definition (\ref{def:pq:V}) and (\ref{GF:V}), 
\begin{equation}
p_k^{(c-1,d-1)}=p_k^{(c,d)}+x     p_{k-1}^{(c,d)}, \quad 
q_k^{(c-1,d-1)}=q_k^{(c,d)}+x^{-1}q_{k-1}^{(c,d)}, \label{cont:1:V}
\end{equation}
\begin{equation}
p_k^{(c,d-1)}=p_k^{(c,d)}-p_{k-1}^{(c,d)}, \quad 
q_k^{(c,d-1)}=q_k^{(c,d)}-q_{k-1}^{(c,d)}, \label{cont:2:V}
\end{equation}
\begin{equation}
\begin{array}{l}
\smallskip
\disp
(k+1)p_{k+1}^{(c,d)}=-(c-d)p_k^{(c,d+1)}-cx     p_k^{(c+1,d+1)}, \\
\disp
(k+1)q_{k+1}^{(c,d)}=-(c-d)q_k^{(c,d+1)}-cx^{-1}q_k^{(c+1,d+1)}. 
\end{array}   \label{cont:3:V}
\end{equation}

Let us prove Lemma \ref{V1}. 
Adding the $(i+1)$-th column multiplied by $x^{-1}$ to the $i$-th column
of $R_{m,n}^{(0,0)}$ for $i=1,2,\ldots,j,~j=m+n-1,m+n-2,\ldots,1$ and
using (\ref{cont:1:V}), we get 
\begin{equation}
R_{m,n}^{(0,0)}=
 \left|
  \begin{array}{ccccc}
   q_1^{(c-m-n+1,d-m-n+1)}      & q_0^{(c-m-n+2,d-m-n+2)}      & \cdots 
 & q_{-m-n+3}^{(c-1,d-1)} & q_{-m-n+2}^{(c,d)} \\
   q_3^{(c-m-n+1,d-m-n+1)}      & q_2^{(c-m-n+2,d-m-n+2)}      & \cdots 
 & q_{-m-n+5}^{(c-1,d-1)} & q_{-m-n+4}^{(c,d)} \\
   \vdots & \vdots & \ddots & \vdots & \vdots \\
   q_{2m-1}^{(c-m-n+1,d-m-n+1)} & q_{2m-2}^{(c-m-n+2,d-m-n+2)} & \cdots 
 & q_{m-n+1}^{(c-1,d-1)}  & q_{m-n}^{(c,d)}    \\
   x^{-m-n+1}p_{2n-1}^{(c-m-n+1,d-m-n+1)} & x^{-m-n+2}p_{2n-1}^{(c-m-n+2,d-m-n+2)}
    & \cdots & x^{-1}p_{2n-1}^{(c-1,d-1)} & p_{2n-1}^{(c,d)} \\
   \vdots & \vdots & \ddots & \vdots & \vdots     \\
   x^{-m-n+1}p_3^{(c-m-n+1,d-m-n+1)}      & x^{-m-n+2}p_3^{(c-m-n+2,d-m-n+2)}
   & \cdots & x^{-1}p_3^{(c-1,d-1)}      & p_3^{(c,d)}      \\
   x^{-m-n+1}p_1^{(c-m-n+1,d-m-n+1)}      & x^{-m-n+2}p_1^{(c-m-n+2,d-m-n+2)}
   & \cdots & x^{-1}p_1^{(c-1,d-1)}      & p_1^{(c,d)}
  \end{array}
 \right|.    \label{shift:1:V}
\end{equation}
Noticing that $p_0=1$ and $p_k=0$ for $k<0$, we see that $R_{m,n}$ can
be rewritten as 
\begin{equation}
R_{m,n}=
 \left|
  \begin{array}{ccccc}
   q_1        & q_0        & \cdots     & q_{-m-n+2} & q_{-m-n+1} \\
   q_3        & q_2        & \cdots     & q_{-m-n+4} & q_{-m-n+3} \\
   \vdots     & \vdots     & \ddots     & \vdots     & \vdots     \\
   q_{2m-1}   & q_{2m-2}   & \cdots     & q_{m-n}    & q_{m-n-1}  \\
   p_{n-m}    & p_{n-m+1}  & \cdots     & p_{2n-1}   & p_{2n}     \\
   \vdots     & \vdots     & \ddots     & \vdots     & \vdots     \\
   p_{-n-m+2} & p_{-n-m+3} & \cdots     & p_1        & p_2        \\
   p_{-n-m}   & p_{-n-m+1} & \cdots     & p_{-1}     & p_0
  \end{array}
 \right|. 
\end{equation}
By the similar calculation to the above, we obtain 
\begin{equation}
R_{m,n}^{(0,0)}=
 \left|
  \begin{array}{ccccc}
   q_1^{(c-m-n,d-m-n)}      & q_0^{(c-m-n+1,d-m-n+1)}      & \cdots 
 & q_{-m-n+2}^{(c-1,d-1)} & q_{-m-n+1}^{(c,d)} \\
   q_3^{(c-m-n,d-m-n)}      & q_2^{(c-m-n+1,d-m-n+1)}      & \cdots 
 & q_{-m-n+4}^{(c-1,d-1)} & q_{-m-n+3}^{(c,d)} \\
   \vdots & \vdots & \ddots & \vdots & \vdots \\
   q_{2m-1}^{(c-m-n,d-m-n)} & q_{2m-2}^{(c-m-n+1,d-m-n+1)} & \cdots 
 & q_{m-n}^{(c-1,d-1)}  & q_{m-n-1}^{(c,d)}    \\
   x^{-m-n}p_{2n}^{(c-m-n,d-m-n)} & x^{-m-n+1}p_{2n}^{(c-m-n+1,d-m-n+1)}
   & \cdots & x^{-1}p_{2n}^{(c-1,d-1)} & p_{2n}^{(c,d)} \\
   \vdots & \vdots & \ddots & \vdots & \vdots     \\
   x^{-m-n}p_2^{(c-m-n,d-m-n)}      & x^{-m-n+1}p_2^{(c-m-n+1,d-m-n+1)}
   & \cdots & x^{-1}p_2^{(c-1,d-1)}      & p_2^{(c,d)}      \\
   x^{-m-n}p_0^{(c-m-n,d-m-n)}      & x^{-m-n+1}p_0^{(c-m-n+1,d-m-n+1)}
   & \cdots & x^{-1}p_0^{(c-1,d-1)}      & p_0^{(c,d)}
  \end{array}
 \right|.    \label{shift:2:V}
\end{equation}
We have from (\ref{cont:1:V}) and (\ref{cont:2:V}) 
\begin{equation}
\begin{array}{l}
\smallskip
\disp
(1+x)p_k^{(c,d)}=p_k^{(c-1,d-1)}+xp_k^{(c,d-1)}, \\
\disp
q_{k+1}^{(c,d-1)}+(1+x^{-1})q_k^{(c,d)}=q_{k+1}^{(c-1,d-1)}. 
\end{array}   \label{cont:a:V}
\end{equation}
Subtracting the $j$-th column multiplied by $(1+x^{-1})^{-1}$ from the $(j+1)$-th column of (\ref{shift:2:V}) for
$j=m+n,m+n-1,\ldots,1$ and using (\ref{cont:a:V}), we get 
\begin{equation}
\begin{array}{l}
\medskip
\disp 
R_{m,n}^{(0,0)}=
(-1)^m(1+x^{-1})^{-m-n}\\
\disp 
\hskip30pt
\times 
 \left|
  \begin{array}{ccccc}
   -q_1^{(c-m-n,d-m-n)}      & q_1^{(c-m-n+1,d-m-n)}      & \cdots 
                             & q_{-m-n+3}^{(c-1,d-2)}     & q_{-m-n+2}^{(c,d-1)}\\
   -q_3^{(c-m-n,d-m-n)}      & q_3^{(c-m-n+1,d-m-n)}      & \cdots 
                             & q_{-m-n+5}^{(c-1,d-2)}     & q_{-m-n+4}^{(c,d-1)}\\
   \vdots & \vdots & \ddots & \vdots & \vdots \\
   -q_{2m-1}^{(c-m-n,d-m-n)} & q_{2m-1}^{(c-m-n+1,d-m-n)} & \cdots 
                             & q_{m-n+1}^{(c-1,d-2)}      & q_{m-n}^{(c,d-1)}   \\
   x^{-m-n}p_{2n}^{(c-m-n,d-m-n)} & x^{-m-n+1}p_{2n}^{(c-m-n+1,d-m-n)} & \cdots 
   & x^{-1}p_{2n}^{(c-1,d-2)}     & p_{2n}^{(c,d-1)} \\
   \vdots & \vdots & \ddots & \vdots & \vdots     \\
   x^{-m-n}p_2^{(c-m-n,d-m-n)}    & x^{-m-n+1}p_2^{(c-m-n+1,d-m-n)}    & \cdots 
   & x^{-1}p_2^{(c-1,d-2)}        & p_2^{(c,d-1)}  \\
   x^{-m-n}p_0^{(c-m-n,d-m-n)}    & x^{-m-n+1}p_0^{(c-m-n+1,d-m-n)}    & \cdots 
   & x^{-1}p_0^{(c-1,d-2)}        & p_0^{(c,d-1)}
  \end{array}
 \right|. 
\end{array}   \label{shift:3:V}
\end{equation}
From (\ref{shift:2:V}) and (\ref{shift:3:V}), we obtain Lemma \ref{V1}. 

Next, we prove Lemma \ref{V2}. 
We have 
\begin{equation}
\begin{array}{l}
\smallskip
\disp 
(k+1)p_{k+1}^{(c,d)}=dp_k^{(c,d+1)}-c(1+x)     p_k^{(c+1,d+2)}, \\
\disp 
(d+k+1)q_{k+1}^{(c,d)}=dq_{k+1}^{(c,d+1)}-c(1+x^{-1})q_k^{(c+1,d+2)}. 
\end{array}   \label{cont:b:V}
\end{equation}
Subtracting the $j$-th column multiplied by 
$\disp \frac{d-m-n+j-2}{(c-m-n+j-1)(1+x^{-1})}$
from the $(j+1)$-th column of (\ref{shift:2:V}) for
$j=m+n,m+n-1,\ldots,1$ and using (\ref{cont:b:V}), we get 
\begin{equation}
\begin{array}{l}
\medskip
\disp 
R_{m,n}^{(0,0)}=
(-1)^{m+n}(1+x)^{-m-n}x^m 
\frac{\disp \prod_{i=1}^m (d-m-n+2i-2)\prod_{k=0}^n(2k+1)}
     {\disp \prod_{j=1}^{m+n}(c-m-n+j-1)} \\
\disp 
\hskip30pt
\times 
 \left|
  \begin{array}{ccccc}
   \widetilde{q}_1^{(c-m-n,d-m-n)}      & q_1^{(c-m-n,d-m-n-1)}      & \cdots 
            & q_{-m-n+3}^{(c-2,d-3)}     & q_{-m-n+2}^{(c-1,d-2)}\\
   \widetilde{q}_3^{(c-m-n,d-m-n)}      & q_3^{(c-m-n,d-m-n-1)}      & \cdots 
            & q_{-m-n+5}^{(c-2,d-3)}     & q_{-m-n+4}^{(c-1,d-2)}\\
   \vdots & \vdots & \ddots & \vdots & \vdots \\
   \widetilde{q}_{2m-1}^{(c-m-n,d-m-n)} & q_{2m-1}^{(c-m-n,d-m-n-1)} & \cdots 
            & q_{m-n+1}^{(c-2,d-3)}      & q_{m-n}^{(c-1,d-2)}   \\
   \widehat{p}_{2n}^{(c-m-n,d-m-n)} & x^{-m-n+1}p_{2n+1}^{(c-m-n,d-m-n-1)} 
    & \cdots                        & x^{-1}p_{2n+1}^{(c-2,d-3)} & p_{2n+1}^{(c-1,d-2)} \\
   \vdots & \vdots & \ddots & \vdots & \vdots     \\
   \widehat{p}_2^{(c-m-n,d-m-n)}    & x^{-m-n+1}p_3^{(c-m-n,d-m-n-1)}
    & \cdots                        & x^{-1}p_3^{(c-2,d-3)}      & p_3^{(c-1,d-2)}      \\
   \widehat{p}_0^{(c-m-n,d-m-n)}    & x^{-m-n+1}p_1^{(c-m-n,d-m-n-1)}
   & \cdots                         & x^{-1}p_1^{(c-2,d-3)}      & p_1^{(c-1,d-2)}
  \end{array}
 \right|. 
\end{array}   \label{shift:4:V}
\end{equation}
Lemma \ref{V2} follows from (\ref{shift:1:V}) and (\ref{shift:4:V}). 

Note that we have 
\begin{equation}
\begin{array}{l}
\smallskip
\disp 
(k+1)p_{k+1}^{(c,d)}=-dxp_k^{(c+1,d+1)}-(c-d)(1+x)p_k^{(c+1,d+2)}, \\
\disp 
(d+k+1)q_{k+1}^{(c,d)}=dq_{k+1}^{(c+1,d+1)}-(c-d)(1+x^{-1})q_k^{(c+1,d+2)}. 
\end{array}   \label{cont:c:V}
\end{equation}
It is easy to see that Lemma \ref{V3} is proved similarly to Lemma
\ref{V2} by using (\ref{cont:2:V}) and (\ref{cont:c:V}). 

The proof of Lemma \ref{V4} is given as follows. 
Adding the $(j-1)$-th column multiplied by $x$ to the $j$-th column of
$R_{m,n}^{(1,1)}$ for $j=m+n,m+n-1,\ldots,2$ and using (\ref{cont:1:V}),
we get 
\begin{equation}
R_{m,n}^{(1,1)}=x^m
 \left|
  \begin{array}{ccccc}
   x^{-1}q_1^+        & q_1        & \cdots     & q_{-m-n+4} & q_{-m-n+3} \\
   x^{-1}q_3^+        & q_3        & \cdots     & q_{-m-n+6} & q_{-m-n+5} \\
   \vdots             & \vdots     & \ddots     & \vdots     & \vdots     \\
   x^{-1}q_{2m-1}^+   & q_{2m-1}   & \cdots     & q_{m-n+2}  & q_{m-n+1}    \\
   p_{n-m}^+    & p_{n-m+1}  & \cdots     & p_{2n-2}   & p_{2n-1}   \\
   \vdots       & \vdots     & \ddots     & \vdots     & \vdots     \\
   p_{-n-m+4}^+ & p_{-n-m+5} & \cdots     & p_2        & p_3        \\
   p_{-n-m+2}^+ & p_{-n-m+3} & \cdots     & p_0        & p_1
  \end{array}
 \right|.     \label{shift:7:V}
\end{equation}
We have from (\ref{cont:1:V}) 
\begin{equation}
\begin{array}{c}
\smallskip
\disp 
p_k^{(c,d)}-x^2   p_{k-2}^{(c,d)}=p_k^{(c-1,d-1)}-x     p_{k-1}^{(c-1,d-1)}, \\
\disp 
q_k^{(c,d)}-x^{-2}q_{k-2}^{(c,d)}=q_k^{(c-1,d-1)}-x^{-1}q_{k-1}^{(c-1,d-1)}. 
\end{array}   \label{cont:d:V}
\end{equation}
Then, subtracting the $(j-1)$-th column multiplied by $x$ from the
$j$-th column of $R_{m,n}^{(-1,-1)}$ for $j=m+n,m+n-1,\ldots,2$ and
using (\ref{cont:d:V}), we get 
\begin{equation}
R_{m,n}^{(-1,-1)}=
 \left|
  \begin{array}{cccc}
   q_1^-      & x^{-1}q_{-1}-xq_1        & \cdots & x^{-1}q_{-m-n+1}-xq_{-m-n+3} \\
   q_3^-      & x^{-1}q_1   -xq_3        & \cdots & x^{-1}q_{-m-n+3}-xq_{-m-n+5} \\
   \vdots     & \vdots                   & \ddots & \vdots                       \\
   q_{2m-1}^- & x^{-1}q_{2m-3}-xq_{2m-1} & \cdots & x^{-1}q_{m-n-1}-xq_{m-n+1}   \\
   p_{n-m}^-    & p_{n-m+1}-x^2p_{n-m-1}   & \cdots & p_{2n-1}-x^2p_{2n-3} \\
   \vdots       & \vdots                   & \ddots & \vdots               \\
   p_{-n-m+4}^- & p_{-n-m+5}-x^2p_{-n-m+3} & \cdots & p_3-x^2p_1           \\
   p_{-n-m+2}^- & p_{-n-m+3}-x^2p_{-n-m+1} & \cdots & p_1-x^2p_{-1}
  \end{array}
 \right|. 
\end{equation}
Noticing that $p_k=q_k=0$ for $k<0$, we obtain 
\begin{equation}
R_{m,n}^{(-1,-1)}=(-x)^m 
 \left|
  \begin{array}{ccccc}
   -x^{-1}q_1^-                              & q_1      & q_0     & \cdots & q_{-m-n+3}\\
   -x^{-1}(q_3^-+x^{-2}q_1^-)                & q_3      & q_2     & \cdots & q_{-m-n+5}\\
   \vdots                                    & \vdots   & \vdots  & \ddots & \vdots    \\
   -x^{-1}(q_{2m-1}^-+\cdots +x^{-2m+2}q_1^-)& q_{2m-1} & q_{2m-2}& \cdots & q_{m-n+1} \\
   p_{n-m}^-+\cdots +x^{2n-2}p_{-n-m+2}^- & p_{n-m+1}  & p_{n-m+2}  & \cdots & p_{2n-1}\\
   \vdots                                 & \vdots     & \vdots     & \ddots & \vdots  \\
   p_{-n-m+4}^-+x^2p_{-n-m+2}^-           & p_{-n-m+5} & p_{-n-m+6} & \cdots & p_3     \\
   p_{-n-m+2}^-                           & p_{-n-m+3} & p_{-n-m+4} & \cdots & p_1
  \end{array}
 \right|.     \label{shift:8:V}
\end{equation}
The first half of Lemma \ref{V4} is obtained by (\ref{shift:7:V}) and
(\ref{shift:8:V}). 
Moreover, we have 
\begin{equation}
D=
 \left|
 \begin{array}{ccccc}
  -x^{-1}q_1^- & x^{-1}q_1-x^{-2}q_0 & q_1-x^{-2}q_{-1} & \cdots & q_{-m-n+4}-x^{-2}q_{-m-n+2} \\
  -x^{-1}q_3^- & x^{-1}q_3-x^{-2}q_2 & q_3-x^{-2}q_1    & \cdots & q_{-m-n+6}-x^{-2}q_{-m-n+4} \\
  \vdots & \vdots & \vdots & \ddots & \vdots \\
  -x^{-1}q_{2m-1}^- & x^{-1}q_{2m-1}-x^{-2}q_{2m-2} & q_{2m-1}-x^{-2}q_{2m-3} & \cdots & q_{m-n+2}-x^{-2}q_{m-n} \\
  p_{n-m+1}^-  & p_{n-m+1}-xp_{n-m}     & p_{n-m+2}-x^2p_{n-m}     & \cdots & p_{2n-1}-x^2p_{2n-3} \\
  \vdots       & \vdots                 & \vdots                   & \ddots & \vdots \\
  p_{-n-m+5}^- & p_{-n-m+5}-xp_{-n-m+4} & p_{-n-m+6}-x^2p_{-n-m+4} & \cdots & p_3-x^2p_1           \\
  p_{-n-m+3}^- & p_{-n-m+3}-xp_{-n-m+2} & p_{-n-m+4}-x^2p_{-n-m+2} & \cdots & p_1-x^2p_{-1}
 \end{array}
 \right|. 
\end{equation}
Subtracting the $2$'nd column from the $1$'st column and using
(\ref{cont:1:V}), we get 
\begin{equation}
\begin{array}{l}
\medskip
\disp 
D=2
 \left|
 \begin{array}{ccccc}
  -x^{-1}q_1 & x^{-1}q_1-x^{-2}q_0 & q_1-x^{-2}q_{-1} & \cdots & q_{-m-n+4}-x^{-2}q_{-m-n+2} \\
  -x^{-1}q_3 & x^{-1}q_3-x^{-2}q_2 & q_3-x^{-2}q_1    & \cdots & q_{-m-n+6}-x^{-2}q_{-m-n+4} \\
  \vdots & \vdots & \vdots & \ddots & \vdots \\
  -x^{-1}q_{2m-1} & x^{-1}q_{2m-1}-x^{-2}q_{2m-2} & q_{2m-1}-x^{-2}q_{2m-3} & \cdots & q_{m-n+2}-x^{-2}q_{m-n} \\
  xp_{n-m}  & p_{n-m+1}-xp_{n-m}     & p_{n-m+2}-x^2p_{n-m}     & \cdots & p_{2n-1}-x^2p_{2n-3} \\
  \vdots       & \vdots                 & \vdots                   & \ddots & \vdots \\
  xp_{-n-m+4} & p_{-n-m+5}-xp_{-n-m+4} & p_{-n-m+6}-x^2p_{-n-m+4} & \cdots & p_3-x^2p_1           \\
  xp_{-n-m+2} & p_{-n-m+3}-xp_{-n-m+2} & p_{-n-m+4}-x^2p_{-n-m+2} & \cdots & p_1-x^2p_{-1}
 \end{array}
 \right|, \\
\disp 
\hskip13pt
=2(-1)^{-m}x^{-2m+1}R_{m,n}^{(0,0)}, 
\end{array}
\end{equation}
which is nothing but the second half of Lemma \ref{V4}. 

From the above discussion, it is easy to verify Lemma \ref{V5}.


\begin{thebibliography}{99}
\bibitem{AK} 
 F. V. Andreev, A. V. Kitaev, 
 Transformations ${RS}_4^2(3)$ of the ranks $\leq4$ and algebraic
 solutions of the sixth Painlev\'e equation, 
 preprint, nlin.SI/0107074.
\bibitem{DM}
 B. Dubrovin, M. Mazzocco, 
 Monodromy of certain Painleve VI transcendents and reflection groups, 
 Invent. Math. {\bf 141} (2000) 55-147.
\bibitem{GtoP}
 K. Iwasaki, H. Kimura, S. Shimomura and M. Yoshida, 
 From Gauss to Painlev\'e -- A Modern Theory of Special Functions, 
 Aspects of Mathematics E16, Vieweg, 1991.
\bibitem{P3:rat}
 K. Kajiwara and T. Masuda, 
 On the Umemura polynomials for the Painlev\'e III equation, 
 Phys. Lett. \textbf{A 260} (1999) 462-467. 
\bibitem{KMNOY}
 K. Kajiwara, T. Masuda, M. Noumi, Y. Ohta and Y. Yamada, 
 Determinant formulas for the Toda and discrete Toda equations, 
 Funkcial. Ekvac. {\bf 44} (2001) 291-307. 
\bibitem{P2:rat}
 K. Kajiwara and Y. Ohta, 
 Determinant structure of the rational solutions for the Painlev\'e II equation,
 J. Math. Phys. {\bf 37} (1996) 4693-4704. 
\bibitem{P4:rat}
 K. Kajiwara and Y. Ohta, 
 Determinant structure of the rational solutions for the Painlev\'e IV equation,
 J. Phys. A: Math. Gen. {\bf 31} (1998) 2431-2446. 
\bibitem{KT1}
 A. N. Kirillov and M. Taneda, 
 Generalized Umemura polynomials, 
 to appear in Rocky Mountain Journal of Mathematics, math.CO/0010279. 
\bibitem{KT2}
 A. N. Kirillov and M. Taneda, 
 Generalized Umemura polynomials and Hirota-Miwa equations, 
 to appear in MSJ Memoirs, math.CO/0106025. 
\bibitem{KT3}
 A. N. Kirillov and M. Taneda, in preparation. 
\bibitem{Koike}
 K. Koike, 
 On the decomposition of tensor products of the representations of the classical groups: 
 by means of the universal characters, 
 Adv. Math. {\bf 74} (1989) 57-86. 
\bibitem{P5:rat}
 T. Masuda, Y. Ohta and K. Kajiwara, 
 A determinant formula for a class of rational solutions of Painlev\'e V equation, 
 to appear in Nagoya Math. J., nlin.SI/0101056. 
\bibitem{Ma}
 M. Mazzocco, 
 Picard and Chazy solutions to the Painleve VI equation, 
 preprint, math.AG/9901054. 
\bibitem{NOOU}
 M. Noumi, S. Okada, K. Okamoto, and H. Umemura,
 Special polynomials associated with the Painleve equations II, 
 In: Saito, M. H., Shimizu, Y., Ueno, K. (eds.) 
 {\it Proceedings of the Taniguchi Symposium, 1997, 
      Integrable Systems and Algebraic Geometry.}
 Singapore: World Scientific, 1998, pp. 349-372. 
\bibitem{NY:P4}
 M. Noumi and Y. Yamada, 
 Symmetries in the fourth Painlev\'e equation and Okamoto polynomials, 
 Nagoya Math. J. {\bf 153} (1999) 53-86. 
\bibitem{NY:P5}
 M. Noumi and Y. Yamada, 
 Umemura polynomials for the Painlev\'e V equation, 
 Phys. Lett. {\bf A247} (1998) 65-69. 
\bibitem{NY1}
 M. Noumi and Y. Yamada, 
 Higher order Painlev\'e equations of type $A_l^{(1)}$, 
 Funkcial. Ekvac. {\bf 41} (1998) 483-503. 
\bibitem{NY2}
 M. Noumi and Y. Yamada, 
 Affine Weyl groups, discrete dynamical systems and Painlev\'e equations, 
 Commun. Math. Phys. {\bf 199} (1998) 281-295. 
\bibitem{NY3}
 M. Noumi and Y. Yamada, 
 A new Lax pair for the sixth Painlev\'e equation associated with 
 $\widehat{\frak{so}}(8)$, 
 preprint. 
\bibitem{private} 
 M. Noumi and Y. Yamada, private communication. 
\bibitem{O1}
 K. Okamoto, 
 Studies on the Painlev\'e equations I, 
 sixth Painlev\'e equation P$_{\rm VI}$, 
 Annali di Matematica pura ed applicata {\bf CXLVI} (1987) 337-381. 
\bibitem{O2}
 K. Okamoto, 
 Studies on the Painlev\'e equations II, 
 fifth Painlev\'e equation P$_{\rm V}$, 
 Japan J. Math. {\bf 13} (1987) 47-76. 
\bibitem{O3}
 K. Okamoto, 
 Studies on the Painlev\'e equations III, 
 second and fourth Painlev\'e equations, P$_{\rm II}$ and P$_{\rm IV}$, 
 Math. Ann. {\bf 275} (1986) 222-254. 
\bibitem{O4}
 K. Okamoto, 
 Studies on the Painlev\'e equations IV, 
 third Painlev\'e equation P$_{\rm III}$,
 Funkcial. Ekvac. {\bf 30} (1987) 305-332. 
\bibitem{Pa}
 P. Painlev\'e, 
 Sur les \'equations diff\'erentielles du second ordre \`a points critiques fixes, 
 C. R. Acad. Sci. Paris {\bf 143} (1906) 1111-1117. 
\bibitem{Pi}
 E. Picard, 
 M\'emoire sur la th\'eorie des functions alg\'ebriques de deux	varables, 
 Journal de Liouville {\bf 5} (1889) 135-319. 
\bibitem{Tane}
 M. Taneda, 
 Polynomials associated with an algebraic solution of the sixth Painlev\'e equation, 
 to appear in Jap. J. Math. {\bf 27} (2002). 
\bibitem{Um}
 H. Umemura, 
 Special polynomials associated with the Painlev\'e equations I,
 preprint. 
\bibitem{YV}
 A. P. Vorob'ev,
 On rational solutions of the second Painlev\'e equation. 
 Diff. Uravn. {\bf 1} (1965) 58-59.
\bibitem{Ya}
 Y. Yamada, 
 Determinant formulas for the $\tau$-functions of the Painlev\'e equations of type $A$, 
 Nagoya Math. J. {\bf 156} (1999) 123-134. 
\end{thebibliography}
\end{document}